\documentclass[aps,twocolumn,superscriptaddress,nofootinbib,amsmath,amssymb,floatfix]{revtex4}

\usepackage{graphicx}
\usepackage{dcolumn}
 \usepackage{hyperref}
\usepackage{bm}
\usepackage{xcolor}
\usepackage{subfigure}
\usepackage{ulem}

\begin{document}

\title{Azimuthal correlations of heavy quarks in Pb+Pb collisions at LHC ($\sqrt{s}=2.76$~TeV)}

\author{Marlene Nahrgang}
\email{nahrgang@subatech.in2p3.fr}
\affiliation{SUBATECH, UMR 6457, Universit\'e de Nantes, Ecole des Mines de Nantes,
IN2P3/CNRS. 4 rue Alfred Kastler, 44307 Nantes cedex 3, France }
\affiliation{Frankfurt Institute for Advanced Studies (FIAS), Ruth-Moufang-Str.~1, 60438 Frankfurt am Main, Germany}

\author{J\"org Aichelin}
\affiliation{SUBATECH, UMR 6457, Universit\'e de Nantes, Ecole des Mines de Nantes,
IN2P3/CNRS. 4 rue Alfred Kastler, 44307 Nantes cedex 3, France }

\author{Pol Bernard Gossiaux}
\affiliation{SUBATECH, UMR 6457, Universit\'e de Nantes, Ecole des Mines de Nantes,
IN2P3/CNRS. 4 rue Alfred Kastler, 44307 Nantes cedex 3, France }

\author{Klaus Werner}
\affiliation{SUBATECH, UMR 6457, Universit\'e de Nantes, Ecole des Mines de Nantes,
IN2P3/CNRS. 4 rue Alfred Kastler, 44307 Nantes cedex 3, France }

\begin{abstract}
In this paper we study the azimuthal correlations of heavy quarks in Pb+Pb collisions with $\sqrt{s}=2.76$~TeV at LHC. Due to the interaction with the medium heavy quarks and antiquarks are deflected from their original direction and the initial correlation of the pair is broadened. We investigate this effect for different transverse momentum classes. Low-momentum heavy-quark pairs lose their leading order back-to-back initial correlation, while a significant residual correlation survives at large momenta.  Due to the larger acquired average deflection from their original directions the azimuthal correlations of heavy-quark pairs are  broadened more efficiently in a purely collisional energy loss mechanism compared to that including radiative corrections. This discriminatory feature survives when next-to-leading-order production processes are included. 
\end{abstract}

\maketitle

\section{Introduction}
Heavy quarks play an important role in the study of the deconfined phase of strongly interacting matter created in ultrarelativistic heavy-ion collisions. Traditional observables of heavy quarks, such as the nuclear modification factor, $R_{\rm AA}$, and the elliptic flow, $v_2$, have intensively been studied at RHIC in the PHENIX \cite{Phenixe} and STAR experiments \cite{Stare,StarD} and in the LHC experiments ALICE \cite{Alice} and CMS \cite{Chatrchyan:2012np}.
All of these data for the $R_{AA}$ signal a significant in-medium energy loss of charm and bottom quarks with larger transverse momenta. The existing data on heavy-quark 
$v_2$ shows that charm quarks partially thermalize within the medium at smaller transverse momentum. 

Both the energy loss of hard probes and the thermalization of the soft part of the heavy-quark spectra result from the interaction of the 
probe with the light partons of the surrounding medium. This interaction can be classified into two  
main contributions, the purely elastic 
cross sections, called  collisional energy loss, \cite{Bjorken,Braaten, Braaten2,Peshier:2006hi,Peigne:2008nd}, and the gluon bremsstrahlung, 
called radiative energy loss \cite{Gyulassy94,Wang95,Baier95,Baier97,Zakharov,GLV,Dokshitzer,AMY,ASW,Zhang04}. 

For light partons, it was originally the collisional energy loss leading to the suppression of high-$p_T$ hadrons which was 
proposed as a signal for the formation of the deconfined quark-gluon plasma (QGP) phase in heavy-ion collisions \cite{Bjorken}. Soon after, the 
radiative energy loss was, however, identified as the dominant energy-loss mechanism, due to its linear increase with the energy $E$ of the
incoming parton in the case of an infinite path length $L$. For a correct description of the radiative energy loss the QCD generalization of 
the Landau-Pomeranchuk-Migdal (LPM) effect  \cite{Baier95,Baier97} needs to be considered, which leads to a reduction of the gluon bremsstrahlung
due to the coherent emission from several scattering centers. As a consequence, the energy loss increases only $\propto \sqrt{E}$ for an infinite path length $L$ and $\propto \hat{q}L^2\ln E$ for a finite path length, where $\hat{q}$ is the jet quenching parameter of the medium.
Further effects due to in-medium modifications of the gluon 
properties \cite{Kampfer00,Djordjevic03,Bluhm:2011sw,Bluhm:2012kp} would reduce the radiative contribution to the energy loss and are 
currently up for debate. The collisional energy loss was better understood through several reconsiderations, including the running of the coupling $\alpha_s$ and calculations beyond 
leading logarithms \cite{Peshier:2006hi,Peigne:2008nd} and its numerical 
importance at intermediate $p_T$ is undoubted. Yet, radiative processes are still commonly thought to be the dominant mechanism for energy loss 
of light partons at large $p_T$.

For heavy quarks, both contributions suffer from mass effects. In standard hard-thermal loop (HTL) calculations \cite{Braaten, Braaten2}, 
the collisional energy loss for relativistic quarks with a finite mass $M$ is essentially an increasing function of the velocity for energies 
$E\ll M^2/T$ and thus presents a mass hierarchy for a given $p_T$. Such a hierarchy is also expected in the radiative energy loss, 
due to the dead cone effect \cite{Dokshitzer}. The relative fraction of radiative and collisional energy loss at a given $p_T$ is therefore 
systematically less under control, but it is suggested that for jets with an energy of the order $\sim 5-15\ \rm{GeV}$ measured in $AA$ collisions, the collisional energy loss might be comparable to 
the radiative one for heavy partons \cite{Mustafa,Ducati,Peshier:2006hi}. Although it remains a challenge to describe $R_{\rm AA}$ and $v_2$ simultaneously in one 
framework, the currently available data on the traditional observables is not sufficient to distinguish well between the two 
different contributions to the energy loss nor between the different models describing these energy-loss mechanisms.

There is a broad range of models which are able to describe $R_{\rm AA}$ or $v_2$ or both by realistic simulations of the heavy-quark
propagation in the medium, usually through rescaling the transport coefficients. These models use purely elastic cross sections \cite{Moore:2004tg,Gossiaux:2008jv,Uphoff:2010sh,Alberico:2011zy,
Uphoff:2011ad}, purely radiative contributions to the energy loss \cite{Abir:2012pu}, a cocktail of both of them including and excluding the LPM effect 
\cite{Gossiaux:2010yx,Gossiaux:2012ya,Cao:2012au}, or nonperturbative approaches, such as resonance scatterings 
\cite{vanHees:2004gq,vanHees:2005wb,vanHees:2007me,Riek:2010fk,Lang:2012cx} or AdS/CFT-based calculations \cite{Horowitz:2007su,Akamatsu:2008,Chesler:2012pw}.

With the improvement of detector and accelerator technologies, new heavy-quark observables may become feasible: e.g., heavy-flavor correlations. In this paper we will investigate the potential of azimuthal correlations between heavy-flavor quark-antiquark pairs ($Q\bar{Q}$) to discriminate between purely collisional and radiative mechanisms. To leading order (LO) in perturbative QCD the production of heavy quarks in initial hard scatterings is given by the processes $q\bar{q}\to Q\bar{Q}$ and  $gg\to Q\bar{Q}$. Due to momentum conservation these 
processes will lead to a back-to-back correlation in azimuthal angle $\Delta \phi$ between the heavy quark and the antiquark. As a consequence of the subsequent interaction with the medium this initial correlation will broaden around $\Delta \phi=\pi$ \cite{Zhu:2006er,Gossiaux:2006yu,Zhu:2007ne,Akamatsu:2009ya,Younus:2013be}. If the heavy quarks thermalized within the medium, the final distribution of heavy-quark-antiquark pairs would be isotropic and the initial correlation would be lost. 
This can be seen by increasing the interaction rates between the heavy quarks and the medium constituents. As final hadronic interactions do not influence the angular correlations the wash-out of initial correlations indicates the presence of a locally thermalized partonic plasma \cite{Zhu:2006er}.
$Q\bar{Q}$ pairs with very small initial momentum are expected to not only lose their initial back-to-back correlation but to be pushed into the same direction by the outward collective flow of the medium and thus obtaining a final correlation around $\Delta\phi\simeq0$. This effect is called ``partonic wind'' \cite{Zhu:2007ne}. Thermalization of heavy quarks with intermediate and large $p_T$ is unlikely, and a residual back-to-back correlation is expected to survive the evolution of the $Q\bar{Q}$ pair in the medium.
At higher center-of-mass beam energies, due to the harder $p_T$ jets, next-to-leading-order (NLO) production processes will become important, which lead to additional initial correlations at $\Delta\phi\simeq 0$. 

The azimuthal correlations of $b$-jet events with large trigger $p_T$ have recently been studied in proton-proton collisions at $\sqrt{s}=7$~TeV in the CMS experiment \cite{CMSbbbar}. They show indeed an enhancement of  correlations in the region of small angular separation as compared to a LO azimuthal back-to-back correlation.

As compared to previous works \cite{Zhu:2006er,Gossiaux:2006yu,Zhu:2007ne,Akamatsu:2009ya,Younus:2013be}, we will systematically investigate the azimuthal correlations between $Q\bar{Q}$ pairs with respect to the two energy-loss mechanisms, collisional and radiative, different $p_T$ trigger classes and LO as well as NLO initializations. We use an improved version of our Monte Carlo approach to the heavy-quark 
propagation in a fluid dynamical medium. In previous publications \cite{Gossiaux:2008jv,Gossiaux:2009mk} it was shown that it 
is able  to reproduce the existing data for $R_{\rm AA}$ and $v_2$ of nonphotonic single electrons at RHIC in a collisional energy-loss scenario by a global rescaling of the rates with a factor $K=2$; predictions for $R_{\rm AA}$ and $v_2$ of D and B mesons at RHIC and LHC were 
provided. By including a first version of radiative energy loss our model turned out to be 
equally able to reproduce the nonphotonic single-electron results at RHIC \cite{Gossiaux:2010yx,Gossiaux:2012hp} after a global rescaling of the rates by $K=0.7$. In \cite{Gossiaux:2012ya}, we have presented a first comparison with the $R_{AA}$ for D mesons
at RHIC and LHC. Although this model was rather successful, one has to mention that these calculations were performed using an outdated $2+1$d fluid dynamical 
medium description of the plasma evolution \cite{Kolb:2003dz}, which relies on an equation of state with a strong first-order phase transition based on ideal hadron and quark-gluon gases. In the present work we use a $3+1$d fluid dynamical description of the medium evolution coming from the initial conditions of the EPOS model  \cite{Werner:2010aa,Werner:2012xh}, which includes an equation of state from lattice QCD calculations \cite{Borsanyi:2010cj}. A first discussion of this new medium description is presented in \cite{Gossiaux:2012ea}. The radiation process off heavy quarks is rectified by implementing a more rigorous phase-space restriction. 
This improved approach is equivalently able to describe the existing heavy-flavor data \cite{Nahrgang:2013xaa}.

This paper is organized as follows: In section \ref{sec:model} we describe the ingredients of the current version of MC@sHQ, 
 coupled to the fluid dynamical evolution. The properties of the collisional and the radiative energy loss mechanisms are presented 
in section \ref{sec:interactions}. Results of our simulations are shown in section \ref{sec:correlations},  for leading-order initialization
in \ref{sec:correlations1} followed by \ref{sec:correlations2}, where we show the additional influence of next-to-leading order production 
processes. The main findings are summarized in the conclusion.

\section{Model}\label{sec:model}

Our approach to the propagation of heavy quarks in the QGP consists of two main parts: the Monte-Carlo implementation of the interaction mechanisms, MC@sHQ, and the EPOS fluid dynamical 
evolution of the QGP medium. While the exhaustive and detailed description of the coupled approach including the comparison to existing data will be given elsewhere, 
we briefly outline the main ingredients, which are of relevance for the present study.

\subsection{EPOS fluid dynamics}
The ideal fluid dynamical background is subsequent to the EPOS initial conditions \cite{Werner:2010aa,Werner:2012xh}, which are obtained from a multiple scattering approach (per nucleon-nucleon collision). Each elementary scattering process is described by a parton ladder, whose final state is a longitudinal color field. The dynamics of this flux tube is described by a relativistic string. In elementary collisions the string breaking by $\bar qq$ production leads to hadron formation from the individual string segments. In nucleus-nucleus collisions the density of flux tubes is large and string segments, which are slow and/or far from the surface, are assumed to quickly constitute locally thermalized matter and then evolve as a fluid. From this procedure one obtains the initial profiles for all fluid dynamical fields. The fluctuating flux tube positions allow us to treat the fluid dynamical evolution event by event accounting for the fluctuating spatial structure of single events. In principle we could thus study the heavy quark propagation in a realistic event-by-event simulation. To gain better statistics, however, we evolve $10^5$ MC@sHQ runs per fluid dynamical event.

The full $3+1$d fluid dynamical simulation is performed including a parametrization of the equation of state from lattice QCD \cite{Borsanyi:2010cj}. It exhibits a crossover transition between partonic and hadronic degrees of freedom in a range of temperatures between $T=145-165$ MeV \cite{Borsanyi:2010bp}.

We use the version EPOS2.17v3 in this work. Here, we try to mimic viscous effects
by taking artificially large values of the flux tube radii (in this case $1$~fm), in order to get smoother initial conditions. This procedure reduces the elliptic flow.

In the present work we concentrate on the azimuthal correlations on the partonic level and do not investigate the effects of hadronization or an interaction of D and B mesons in a final hadronic stage. All results presented in the following are shown for charm and bottom quarks taken locally at a transition temperature of $T_{\rm c}=155$~MeV, which is well within the range given by lattice QCD.

 In its integral version including a final hadronic cascade the EPOS approach is able to simultaneously describe a variety of soft observables such as particle yields, spectra, flow coefficients, and dihadron correlations at RHIC and LHC energies  \cite{Werner:2010aa,Werner:2012xh}. Having the soft sector under control gives us confidence that we can reliably investigate the in-medium modifications of the  heavy-quark distributions.

\subsection{MC@sHQ}
The Monte Carlo sampling of the scatterings of the charm ($m_c=1.5$~GeV) and bottom ($m_b=5.1$~GeV) quarks with the light partons and gluons is performed by solving the Boltzmann equation with the respective cross sections for collisional and radiative processes. Locally the temperature and the fluid velocity are taken from the fluid dynamical evolution. They provide the local thermal distribution of the medium constituents and enter into the reaction rates.  In this work, the light quarks enter the thermal distribution as massless, relativistic particles.

\subsubsection{Initialization}
Understanding the measurements of charm and bottom production in hadronic collisions at the Fermilab Tevatron and the LHC has largely evolved by introducing the fixed-order next-to-leading-log (FONLL) framework \cite{FONLL1,FONLL2,FONLL3}.
It combines fixed LO QCD with a resummation to all orders with next-to-leading log accuracy. Theoretical uncertainties are well under control and can be estimated from variations of the factorization and renormalization scale, the heavy-quark mass, and uncertainties in the parton distribution functions.
In the standard version of MC@sHQ we use the initial $p_T$ distribution from FONLL and assume a flat rapidity distribution in the range of $y=[-1,1]$, 
with nucleon-nucleon cross sections of 
$\left.\frac{d\sigma_{c}}{dy}\right|_{y=0}=\left.\frac{d\sigma_{\bar{c}}}{dy}\right|_{y=0}=769~\mu{\rm b}$ and 
$\left.\frac{d\sigma_{b}}{dy}\right|_{y=0}=\left.\frac{d\sigma_{\bar{b}}}{dy}\right|_{y=0}=25~\mu{\rm b}$ at $\sqrt{s}=2.76~{\rm TeV}$ \cite{fonllform}. 
The FONLL framework allows us to calculate one-particle inclusive distributions only, because the information of the rest of the event is integrated over. We, thus, consider initial azimuthal correlations only to LO processes, which give a delta peak at $\Delta\phi=\pi$. 
For a study of primarily theoretical interest, we will use this initialization in section \ref{sec:correlations1}. A more realistic initial situation is considered in section \ref{sec:correlations2}, where a Monte Carlo implementation to NLO QCD matrix elements plus parton shower evolution of the initial and final state, MC@NLO \cite{Frixione:2003ei,Frixione:2002ik}, is used to generate the initial $Q\bar{Q}$ pairs event-by-event.

In the coupled approach of MC@sHQ+EPOS the $Q\bar{Q}$ pairs are initialized randomly over the spatial points of initial  nucleon-nucleon scatterings. 
During the pre-equilibrium stage, until the fluid dynamical evolution starts at $\tau_0=0.35$~fm, the heavy quarks do not undergo any scatterings but are evolved via free streaming.

\subsubsection{Collisional energy loss}

The rate for the elastic $2\rightarrow 2$ collisional processes $Q+q\rightarrow Q'+q'$ and 
 $Q+g\rightarrow Q'+g'$ in a fluid cell at rest is generically written as 
\begin{equation}
R_i = \frac{1}{E}
     \int \frac{{\rm d}^3k}{(2\pi)^3} n_i(k) \frac{p\cdot k}{k^0}
     \int {\rm d}t \frac{{\rm d}\sigma_{i,2\to 2}}{{\rm d}t}
\label{eq:rate}
\end{equation}
where $p$ and $E=p_0$ are the four--momentum and the energy of the incoming heavy quark, respectively, 
 and $k$ is the four--momentum of the incoming light quark or gluon. $n_i(k)$ is the thermal distribution 
of the light quarks ($i=q$) or gluons ($i=g$), which is taken as of Boltzmann type in the current version of MC@sHQ used for the present study, and ${\rm d}\sigma_{i,2\rightarrow 2}/{\rm d}t$ is the differential cross section averaged (summed) on entrance (exit) polarizations and colors, calculated using matrix elements ${\cal M}_i$ as 
\begin{equation}
\frac{{\rm d}\sigma_{i,2\rightarrow 2}}{{\rm d}t}=\frac{1}{64\pi s}\frac{1}{|\vec{p}_{\rm cm}|^2}|{\cal M}_{i,2\to 2}|^2\,,
\label{eq:elcrosssection}
\end{equation} 
where 
the matrix elements for the various channels are calculated from the (regularized) pQCD Born
approximation~\cite{Svetitsky:1987gq,Combridge:1978kx}. In MC@sHQ, first $\vec{k}$ is sampled
according to the weight $n_i(k) p\cdot k/k^0 \sigma_{i,2\rightarrow 2}(s)$ with $s$ being the Mandelstam variable; next, the $t$
Mandelstam variable is sampled according to $ {\rm d}\sigma_{i,2\rightarrow 2}/{\rm d}t$. 
For the $t$ channel the matrix elements need however to be regularized in the infrared \cite{Weldon:1982aq}. 
In our approach the matrix elements are evaluated according to the following theoretical considerations:

\begin{itemize}
\item[\textbullet] HTL+semihard: In media at finite temperature the Born approximation to the scattering matrix element is not justified for low momentum 
transfers $|t|$ \cite{Braaten, Braaten2}.  Here, collective modes of the medium dominate and gluon propagators need to be resummed by using the hard-thermal loop (HTL) approach.
At large $|t|$, however, these collective phenomena are unimportant and the bare gluon propagator can be used.
The average energy loss can thus be obtained from an approach that combines HTL at low $|t|$ and hard calculations at large $|t|$. 
In a weak-coupling theory such as QED, the final result  is independent of the intermediate scale $t^*$ 
separating the low and the large $|t|$ scale \cite{Braaten, Braaten2}.
In QCD, however, the underlying condition that $m_D^2\ll T^2$ is violated at temperatures reached in heavy-ion collisions. The HTL+hard approach does, thus, explicitly depend on $t^*$.
Physically, this means that the screening distance is of the same order as the average distance of medium constituents 
and that hard processes are also influenced by polarizations of the medium. By adding a gluon self energy to the hard gluon propagator (and calling this approach 
HTL+semihard) we were able to resolve this situation and obtain an average energy loss independent of the intermediate scale $t^*$.
For integration into our model, we then evaluate the differential cross sections on the full $|t|$ range using an effective gluon propagator with a self-energy calibrated to reproduce that energy loss, namely
\begin{equation}
\frac{1}{t}\rightarrow \frac{1}{t-0.2\tilde{m}_D^2(T)}\,,
\end{equation}
with a self-consistent Debye mass evaluated as (see \cite{Gossiaux:2008jv} for details)
\begin{equation}
\tilde{m}_D^2(T)= \frac{N_c}{3} \left(1+\frac{n_f}{6}\right)4\pi\,\alpha_s[-\tilde{m}_D^2(T)]\,T^2\,.
\end{equation}
\item[\textbullet] Running $\alpha_s$: 
It was shown that the failure of fixed-coupling pQCD calculations to give the correct (as compared to lattice QCD calculations) Debye mass $m_D$ can be remedied by properly taking 
the running of the strong coupling constant $\alpha_s$ into account \cite{Peshier:2006hi}. In this way the Debye mass is calculated self-consistently from $\alpha_s$ at the scale 
of the Debye mass itself. 
By following this procedure the average collisional energy loss is claimed \cite{Peshier:2006hi} to be more important than in
the fixed-$\alpha_s$ calculation \cite{Braaten2}. In \cite{Peigne:2008nd}, the running of $\alpha_s$ is rigorously implemented. The average energy loss is, however, not found to be larger than in  \cite{Braaten2} in the region where the calculations are applicable, i.e., for large momentum $p$ of the incoming parton and large temperature $T$ of the medium.
As explained in \cite{Gossiaux:2008jv}, we use a phenomenological parametrization of the running $\alpha_s$ extracted from experimental data \cite{Mattingly:1993ej,Brodsky:2002nb} 
and constrained by theory~\cite{Dokshitzer:1995qm} that is infrared finite. Extending our HTL+semihard prescription to the running $\alpha_s$ case, one obtains values of energy 
loss which are in good agreement with the results in \cite{Peigne:2008nd} for large momenta $p$ and temperatures $T$ and which exceed the average energy loss as in \cite{Braaten2} by a factor of $\sim 2$ for intermediate momenta $p$ and temperatures $T$.
 \end{itemize}

Due to theoretical uncertainties in the perturbative calculations, the obtained cross sections need to be compared to a reference. 
Naturally, one would wish to scale the associated transport coefficients to lattice QCD calculations. Unfortunately, they cannot yet be precisely 
and reliably calculated within lattice QCD. One is thus compelled to compare final results of $R_{\rm AA}$ to available experimental data. As is common for models including only collisional energy loss (see, e.g., \cite{Molnar:2004}), the scattering rates need 
to be rescaled by a global factor (here named $K$) larger than unity in order 
to be able to reproduce the data for $R_{\rm AA}$. In realistic simulations, however, besides the necessity to 
include energy loss by gluon bremsstrahlung, further effects such as the modeling of the medium expansion\footnote{In particular, different descriptions of the medium evolution can 
lead to factors as large as $2$ for the $R_{\rm AA}$ at large $p_T$ \cite{Gossiaux:2011ea}.}, the strength of the coupling constant\footnote{We choose 
$\alpha_s(Q^2=0)\approx 1.2$, which is small as compared to the fit proposed in \cite{Deur:2008} for which $\alpha_s(Q^2=0)\approx \pi$.}, initial state cold nuclear matter 
effects and hadronization play a role and affect the precise value of the $K$-factor.  With $K=1.5$ our model of purely collisional energy loss is able to describe not only $R_{\rm AA}$ of D and B mesons and of heavy-flavor electrons reasonably well 
but also the elliptic flow $v_2$ at LHC \cite{inpreparation} leaving the possibility of nuclear shadowing at lower $p_T$ and a contribution to $v_2$ from a possible hadronic phase.

\subsubsection{Radiative energy loss}
As mentioned in the introduction, several calculations of radiative energy loss can be found in the literature 
for the case of a massless parton, 
and some of them have been extended to the case of a heavy quark \cite{Dokshitzer, ASW, Zhang04, Djordjevic05,Zakharov04}.
Usually, those approaches rely on the eikonal limit for which the formation time of the radiated gluon is large with respect to the mean-free path. This implies that several collisions
with partons of the medium contribute coherently to the radiation of a single gluon, which leads to the LPM-type suppression. In \cite{Gossiaux:2010yx,Aichelin:2013mra}, we adopt a 
different viewpoint: Since the mass of the heavy quarks acts as a regulator of the collinear divergence the formation time is reduced. We extend the calculations \cite{Gunion:1981qs} for incoherent radiation off a single massless parton to the case of massive quarks. In \cite{Aichelin:2013mra}, it is shown that differential cross section for 
the $Q+q\rightarrow Q'+q'+g$ and $Q+g\rightarrow Q+g'+g''$ radiative processes can be written as
\begin{equation}
\frac{{\rm d}\sigma^{Qq\rightarrow Qgq}}{{\rm d}x {\rm d}^2k_t {\rm d}^2l_t} =
\frac{1}{2(s-m_Q^2)}|\mathcal{M}_{2\to 3}|^2 \frac{1}{4(2\pi)^5\,\sqrt{\Delta}}
\Theta(\Delta)\; ,
\label{eq:xsec}
\end{equation}
where $m_Q$ is the mass of the heavy quark, $l_t$ is the momentum transfer -- essentially of 
transverse nature -- induced by the light parton, $x$ is the momentum fraction carried away by the radiated
 gluon, and $k_t$ is its transverse momentum, while 
\begin{multline}
 \Delta=\left(x(1-x)\,s-x\,m_Q^2-k_t^2+2x\,\vec{k}_t\cdot
 \vec{l}_t\right)^2-\\4x(1-x)\,l_t^2\,(x\,s-k_t^2)
\end{multline}
 is associated with the measure in phase space. 
Special emphasis is put on the exact conservation of energy and momentum through the 
$\Theta(\Delta)$ condition, which plays a crucial role for heavy quarks at intermediate momenta.
While the exact expression for $|{\cal M}|^2$ at finite energy is too cumbersome 
\cite{Kunszt:1980} to be implemented in a Monte Carlo generator, it was shown in \cite{Aichelin:2013mra} --
 as well as in \cite{fochler:2013} for the case of massless quarks -- that a fair agreement with the exact
  calculation can be achieved for quantities such as  $x{\rm d}\sigma/{\rm d}x$ and the average energy loss by considering the eikonal limit in $|{\cal M}|^2$ but preserving the phase-space condition 
  $\Theta(\Delta)$. In \cite{Aichelin:2013mra}, we then propose to approximate
\begin{equation}
  \frac{{\rm d}\sigma^{Qq\rightarrow Qgq}}{{\rm d}x {\rm d}^2k_t {\rm d}^2l_t}\simeq
  \frac{1}{\pi}\frac{{\rm d}\sigma_{\rm el}}{{\rm d}t}\, P_g(x,\vec{k}_t,\vec{l}_t)\Theta(\Delta)
  \label{eq:model2}
  \end{equation}
for the so-called "QCD" gauge-invariant contribution dominating the radiation
spectrum in a $Q+q\rightarrow Q'+q'+g$ process. In equation~(\ref{eq:model2}), ${\rm d}\sigma_{\rm el}/{\rm d}t$ is
the differential  $Q+q\rightarrow Q'+q'$ cross section defined in equation~(\ref{eq:elcrosssection}), taken with
$t=l_t^2$, and
\begin{multline}
P_g(x,\vec{k}_t,\vec{l}_t)=\frac{3\alpha_s}{\pi^2}\frac{1-x}{x}
  \bigg(\frac{\vec{k}_t}{k_t^2+x^2m_Q^2}
  -\\
  \frac{\vec{k}_t-\vec{l}_t}{(\vec{k}_t-\vec{l}_t)^2 +x^2 m_Q^2}\bigg)^2
  \label{eq:gluon_distribution}
  \end{multline}
is an extra radiation factor similar to the one found in~\cite{Gunion:1981qs}
for massless quarks. Thanks to the dominance of the $t$ channel, the $Q+g\rightarrow Q+g'+g''$ process 
can then be modeled using equation~(\ref{eq:model2}) with ${\rm d}\sigma_{\rm el}/{\rm d}t$ taken as the 
$Q+g\rightarrow Q+g'$ differential cross section. In the Monte Carlo generator, an explicit realization of 
the elastic process is achieved first, and the radiation factor $P_g$ is then sampled along the variables 
$x$ and $\vec{k}_t$.

In \cite{Gossiaux:2012hp},the implementation of radiative processes was generalized to include 
the effect of coherence -- hereafter referred to as ``radiative + LPM''. For this purpose, the
$N_{\rm coh}$ coherent collisions with light partons responsible for the radiation of a single gluon were modeled by 
an effective scattering center, and a quenching factor was deduced for the power spectrum per unit length
${\rm d}^2I/({\rm d}z {\rm d}\omega)$ as compared to the incoherent radiation. In our Monte Carlo procedure, this quenching factor
is thus systematically applied for $2\rightarrow 3$ radiative processes, in order to account for the coherence effects. 
It turns out, however, that these effects are of minor practical importance 
for the actual values of the D-meson $R_{AA}$ in the $p_T$ range of $0-20$~GeV considered in this work.

For a combination of collisional and radiative (+ LPM) energy loss, we need to rescale the scattering rates by a global factor $K=0.8$ in order to reproduce the available data 
(for D and B mesons and heavy-flavor electrons) for $R_{\rm AA}$ and $v_2$ at LHC with the same conclusions as for the purely collisional scenario mentioned above \cite{Nahrgang:2013xaa,inpreparation}. 

 \begin{figure}[tb]
   \subfigure{\label{fig:disperp5}\includegraphics[width=0.48\textwidth]{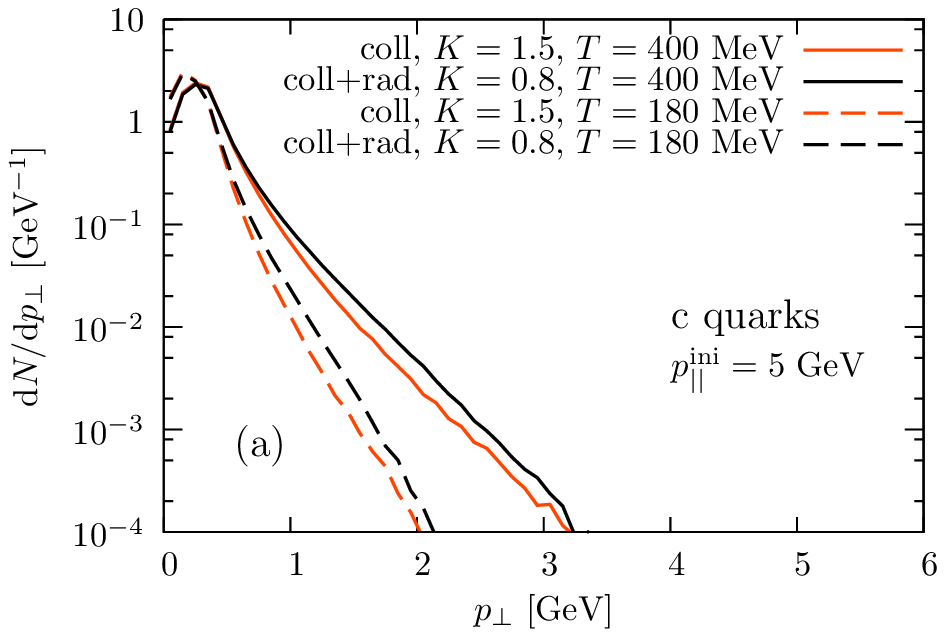}}
 
   \subfigure{\label{fig:disperp25}\includegraphics[width=0.48\textwidth]{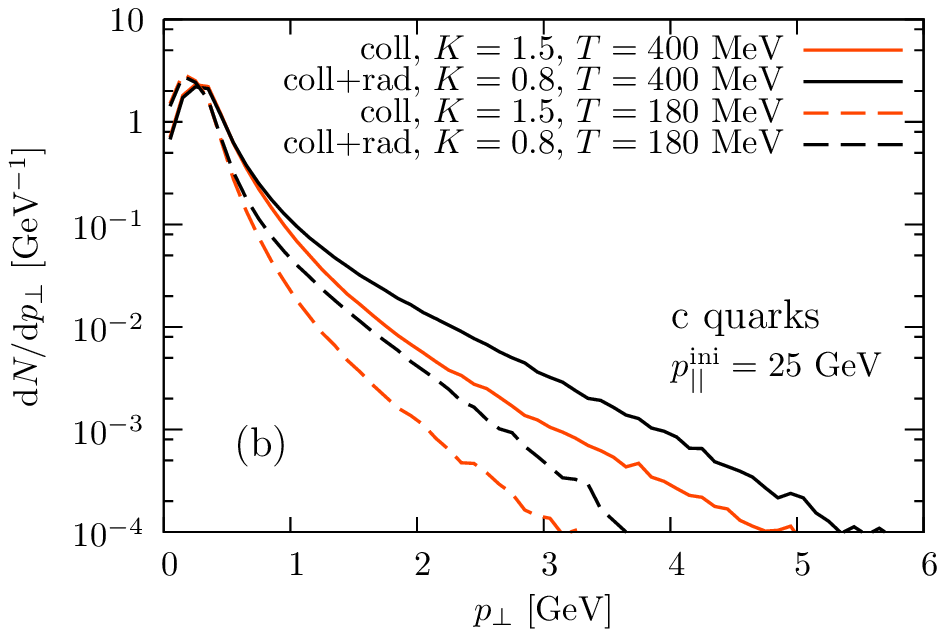}}

 \caption{(Color online) The distribution of $p_\perp$ that a charm quark with $p_{||}^{\rm ini}=5$  \subref{fig:disperp5} and $p_{||}^{\rm ini}=25$  \subref{fig:disperp25} acquires in one scattering, purely collisional (orange/light) or including radiative 
corrections (black/dark) with medium constituents at temperature $T=400$~MeV (solid) and $T=180$~MeV (dashed).}
 \label{fig:dispperp}
 \end{figure}

 \begin{figure}[tb]
   \subfigure{\label{fig:Nscatc}\includegraphics[width=0.48\textwidth]{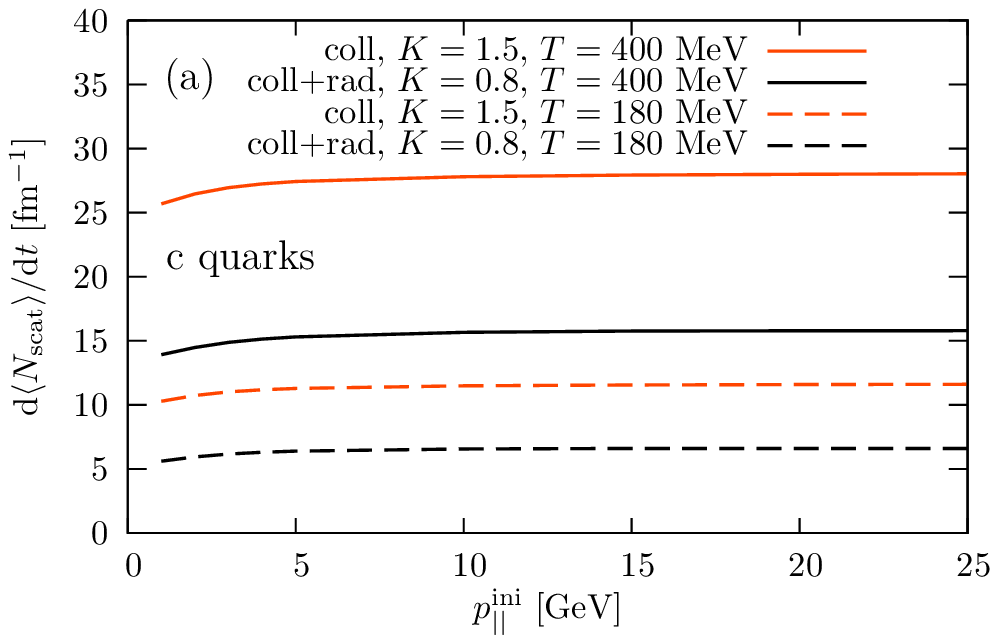}}
 
   \subfigure{\label{fig:Nscatb}\includegraphics[width=0.48\textwidth]{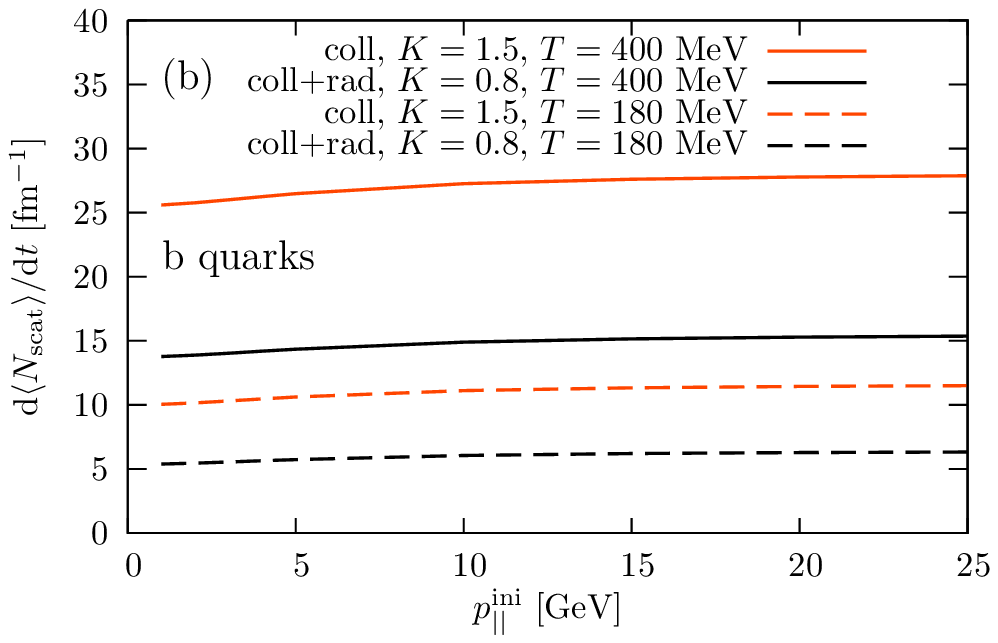}}

 \caption{(Color online) Scattering rate of a charm \subref{fig:Nscatc} and a bottom  \subref{fig:Nscatb} quark at an initial momentum $p_{||}^{\rm ini}$ with medium constituents for $T=400$~MeV (solid) and $T=180$~MeV (dashed). We compare the purely collisional scatterings (orange/light) with those including radiative corrections (black/dark).}
 \label{fig:Nscat}
 \end{figure}

\section{Properties of the interactions}\label{sec:interactions}

In what follows, the two types of interactions, which were outlined in the previous section, have to be understood in connection with their corresponding $K$--factor, i.~e., $K=1.5$ or $K=0.8$, when we speak of ``collisional'' or ``collisional and radiative'' respectively. Moreover, when referring to radiative energy loss we implicitly mean that the LPM suppression is also included.

In this section we will first present exact results for an infinitesimal time step and then toward the end discuss the effect of a small but finite evolution time of the heavy quarks in a static, infinite medium at a given temperature. This is a useful step toward the full coupling to a fluid dynamically expanding medium.

It is instructive to first analyze the basic properties of the interactions between a heavy quark and a light parton from a thermal medium at a given temperature. In this setup the heavy quark has an initial momentum $\vec{p}^{\,\rm ini}=(0,0,p_{||}^{\rm ini})$. 
The evolution of azimuthal correlations in the medium is determined by how effectively the heavy (anti)quarks acquire momentum perpendicular to their original directions determined by $\vec{p}^{\,\rm ini}$. With the given initialization this quantity is thus $p_\perp=\sqrt{p_1^2+p_2^2}$.

A first basic property of the interaction is the scattering rate, which for elastic $2\to 2$ processes corresponds to the expression in equation~(\ref{eq:rate}):
\begin{multline}
\frac{{\rm d} N_{\rm scat}}{{\rm d}t} = \frac{1}{2E}
     \sum_{i=q,g}\int \frac{n_i(k) {\rm d}^3k}{(2\pi)^3 2k_0}  
     \int (2\pi)^4 |{\cal M}_{i,2\rightarrow n}|^2\\
     {\rm d}\Phi_n(p+k;p'\cdots p_n)\; ,
     \label{def:rateinvariant}
\end{multline}
where $n=2,3$ for elastic and radiative processes, respectively, $p'$ is the four-momentum of the 
outgoing heavy quark, and $d\Phi_n$ is the usual invariant phase space of the exit channel,
\begin{equation}
{\rm d}\Phi_n(P;p_1\cdots p_n) =\delta^4\left(P-\sum_{j=1}^n p_j\right) \prod_{j=1}^n \frac{{\rm d}^3p_j}{(2\pi)^3 2E_j}\,.
\end{equation}
From this, one can build more differential observables such as the rate of deflection from $p_\perp^{\,\rm ini}=0$ toward finite $p_\perp$, defined as
\begin{multline}
\frac{{\rm d} N_{\rm scat}}{{\rm d}t {\rm d}p_\perp} = \frac{1}{2E}
     \sum_{i=q,g}\int \frac{n_i(k) {\rm d}^3k}{(2\pi)^3 2k_0}  
     \int (2\pi)^4 |{\cal M}_{i,2\to n}|^2\\
     {\rm d}\Phi_n(p+k;p'\cdots p_n) \delta(p'_\perp-p_\perp)\; ,
     \label{eq:difrate}
\end{multline}
which trivially satisfies $\int {\rm d}p_\perp {\rm d} N_{\rm scat}/({\rm d}t {\rm d}p_\perp)=
{\rm d} N_{\rm scat}/{\rm d}t$. Technically, one generates $2\to n$ processes with our Monte Carlo routines and bins the outgoing heavy quarks in $p_\perp$. The routines used for evaluating the properties of the interactions in this setup are identical to the ones used for the full evolution in the next sections. 

 Figure \ref{fig:dispperp} shows the $p_\perp$ distribution in a single scattering of charm quarks with the medium constituents for two different initial parallel momenta, $p_{||}^{\rm ini}$, and two different temperatures. This quantity is defined as the ratio $\frac{{\rm d} N_{\rm scat}}{{\rm d}t {\rm d}p_\perp}/\frac{{\rm d} N_{\rm scat}}{{\rm d}t}$ and is thus  normalized with respect to the integration over $p_\perp$. 
  We see that this distribution extends to higher $p_\perp$ for increasing both the initial parallel momentum and the temperature of the medium. It is also evident that the average $p_\perp$ acquired in one purely elastic collision is smaller than that in a scattering with radiative corrections. Although the distribution of $p_\perp$ in a single scattering does not depend on the $K$-factor we mention it already here to better outline the following arguments. The $K$-factor crucially affects the scattering rate, which can simply be obtained from equation (\ref{def:rateinvariant}) and is  
 shown in figure \ref{fig:Nscat}. Note here that the absolute number of radiative processes diverges due to the divergence of soft gluon emission. We, thus, apply a minimum fraction of longitudinal momentum $x=0.05$ of the emitted gluons with respect to the emitting heavy quark. The rates practically do not depend on $p_{||}^{\rm ini}$ but decrease strongly with  temperature. The most important aspect for our considerations is that the scattering rate for the purely collisional interaction is larger than for the combined (collisional plus radiative corrections) interaction.

\begin{figure}[tb]
   \subfigure{\label{fig:timedisperp5}\includegraphics[width=0.48\textwidth]{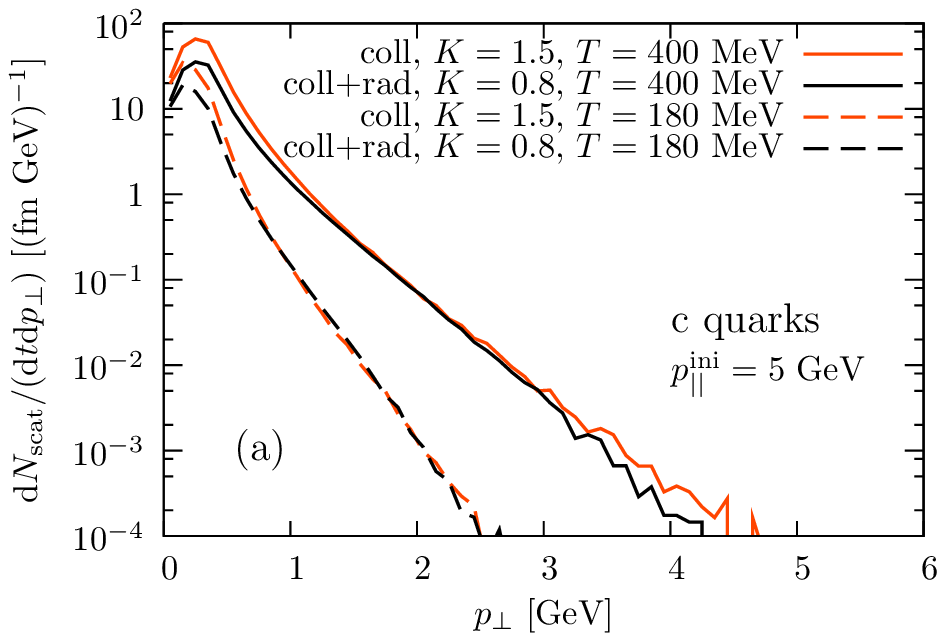}}
 
   \subfigure{\label{fig:timedisperp25}\includegraphics[width=0.48\textwidth]{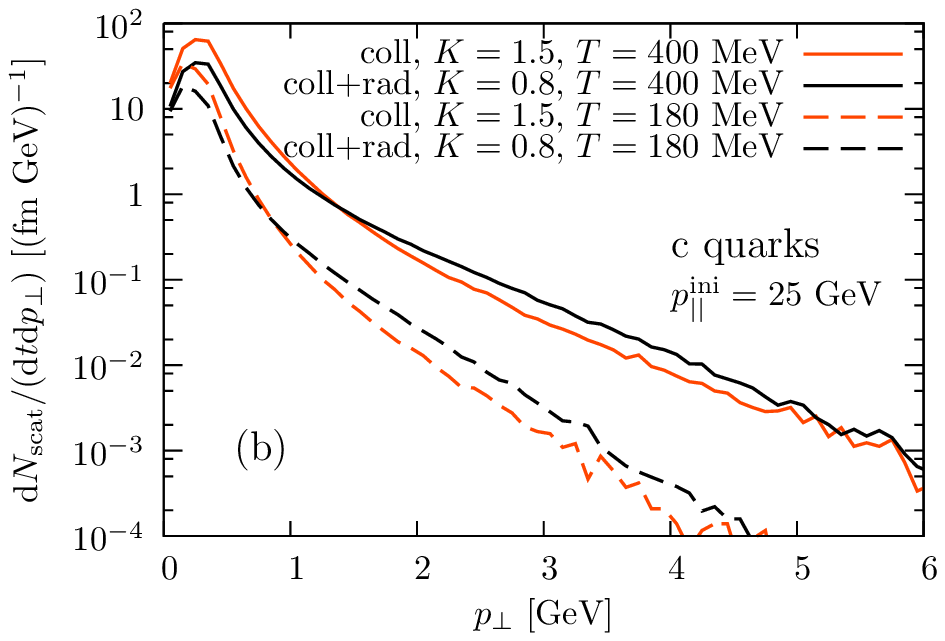}}

 \caption{(Color online) The product of the distribution in figure \ref{fig:dispperp} and the scattering rate in figure \ref{fig:Nscat} for the respective initial momenta of the charm quark, $p_{||}^{\rm ini}=5$  \subref{fig:disperp5} and $p_{||}^{\rm ini}=25$  \subref{fig:disperp25} in the purely collisional scenario (orange/light) or including radiative corrections (black/dark). The medium constituents have temperature $T=400$~MeV (solid) and $T=180$~MeV (dashed).}
 \label{fig:timedispperp}
 \end{figure}

Next, in figure \ref{fig:timedispperp}, we investigate 
the distribution defined by 
equation~(\ref{eq:difrate}) for charm quarks with initial parallel momenta $p_{||}^{\rm ini}=5$ and $p_{||}^{\rm ini}=25$ undergoing scatterings with thermal medium constituents at two different temperatures via the two interaction mechanisms. It is the product of the quantities shown in figures \ref{fig:dispperp} and \ref{fig:Nscat}. Due to the higher scattering rate for the purely collisional interaction, we see that the differences between the two interaction mechanisms become smaller. For small initial momenta one even observes that the rate of deflection toward finite $p_\perp$ is larger for the purely collisional interaction mechanism.

\begin{figure}[tb]
   \subfigure{\label{fig:pperp2c}\includegraphics[width=0.48\textwidth]{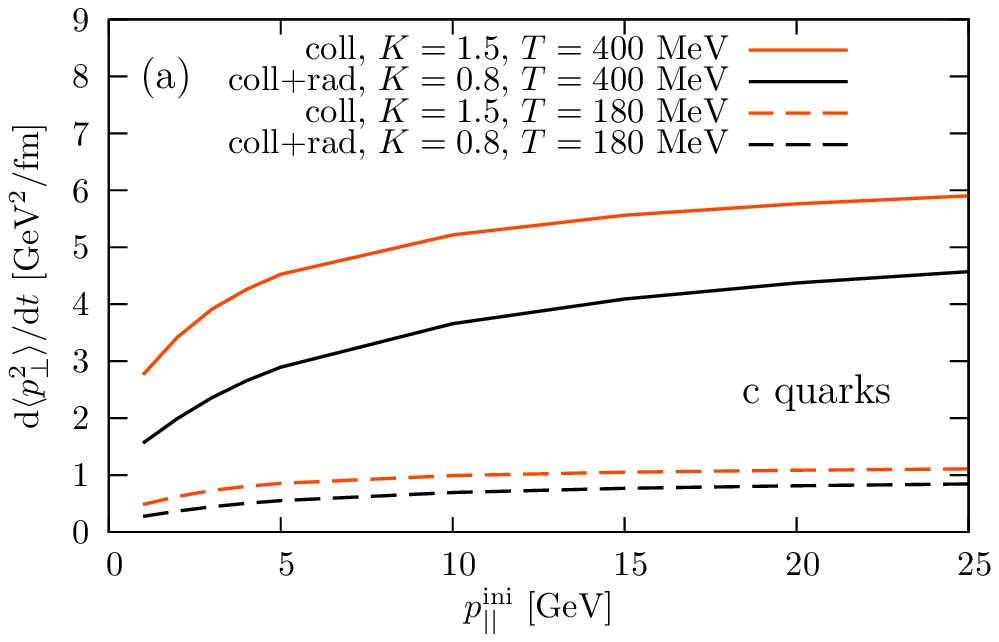}}
 
   \subfigure{\label{fig:pperp2b}\includegraphics[width=0.48\textwidth]{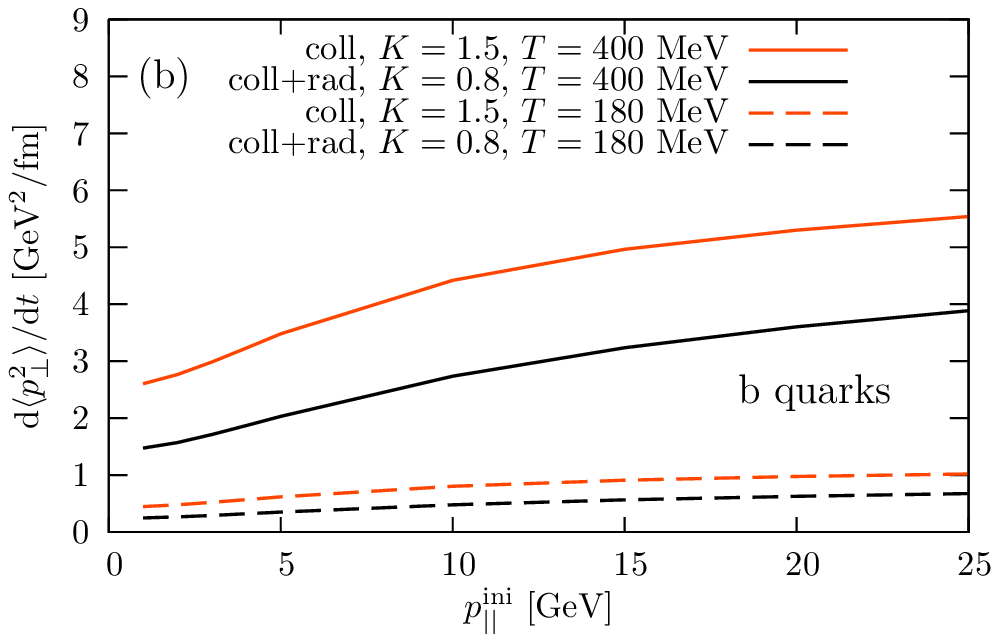}}

 \caption{(Color online) Average  $p_\perp^2$ acquired by a charm \subref{fig:pperp2c} and bottom  \subref{fig:pperp2b} quark with $p_{||}^{\rm ini}$ interacting with medium constituents at $T=400$~MeV (solid) and $T=180$~MeV (dashed). We compare the purely collisional scatterings (orange/light) with including radiative corrections (black/dark).}
 \label{fig:pperp2}
 \end{figure}

 We now proceed to the study of the rate by which the heavy quark with $p_{||}^{\rm ini}$ acquires an average $p_\perp^2$  in figure \ref{fig:pperp2}. Especially at higher temperatures the clear difference between the two basic interaction mechanisms is well reflected. For all initial momenta the purely collisional scatterings lead to a larger average $p_\perp^2$ than that obtained including the radiative corrections. The same ratio of the average $p_\perp^2$ of the two interaction mechanisms is observed at a smaller temperature, where, however, the average $p_\perp^2$ is much smaller and the absolute difference between the two types of interactions is less pronounced. It is also interesting to note that the average $p_\perp^2$ is mostly flavor independent. In \cite{Nahrgang:2013pka} we investigated the same quantity for a purely radiative energy loss mechanism, where $K=1.8$. The average $p_\perp^2$ in this case is even smaller than that for the mechanisms investigated here.

 \begin{figure}[tb]
   \subfigure{\label{fig:dragc}\includegraphics[width=0.48\textwidth]{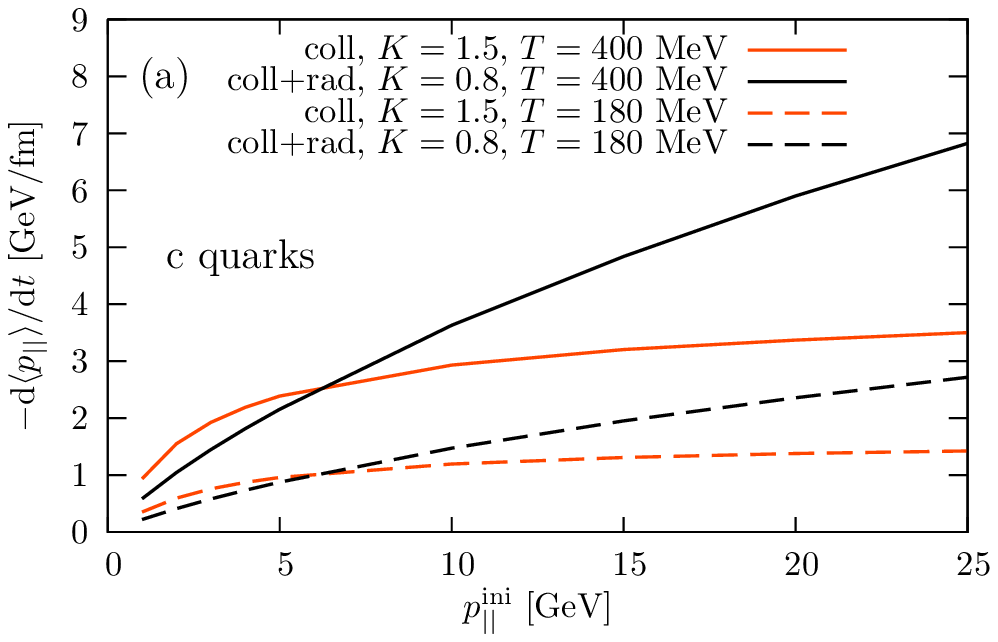}}
 
   \subfigure{\label{fig:dragb}\includegraphics[width=0.48\textwidth]{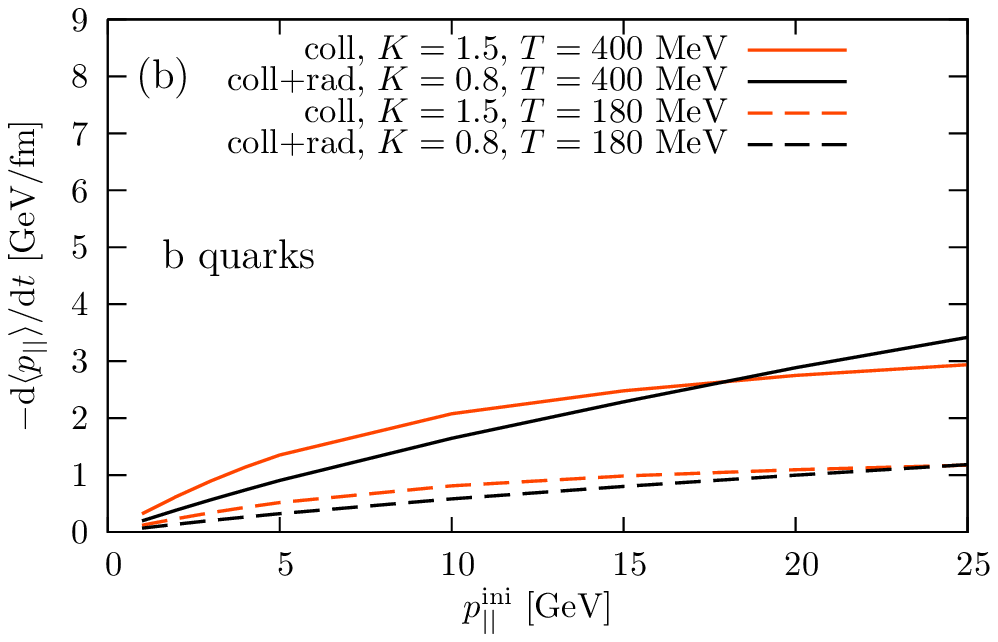}}

 \caption{(Color online) The drag coefficient of a charm \subref{fig:dragc} and a bottom  \subref{fig:dragb} quark with $p_{||}^{\rm ini}$ interacting with medium constituents at $T=400$~MeV (solid) and $T=180$~MeV (dashed). We compare the purely collisional scatterings (orange/light) with including radiative corrections (black/dark).}
 \label{fig:drag}
 \end{figure}

For a full evolution through the medium, the final $p_\perp$ does indirectly also depend on the drag coefficient, which is defined as the rate of losing parallel momentum. When the heavy quark quickly loses $p_{||}$ the collisions will be less effective in transferring a large $p_\perp$. We analyze the drag coefficient in figure \ref{fig:drag}. We observe the nearly linear increase with $p_{||}^{\rm ini}$ for the interaction scenario including radiative corrections and a slower increase for the purely collisional one at large $p_{||}$. In the given range of $p_{||}^{\rm ini}$, the drag coefficient for charm quarks shows a strong dependence on the temperature of the medium. For higher $p_{||}^{\rm ini}$ it will show a similar temperature dependence for bottom quarks as well. The drag coefficient is larger for charm quarks, figure \ref{fig:dragc}, than for bottom quarks, \ref{fig:dragb}. 

 \begin{figure}[tb]
   \subfigure{\label{fig:evoltimedisperp5}\includegraphics[width=0.48\textwidth]{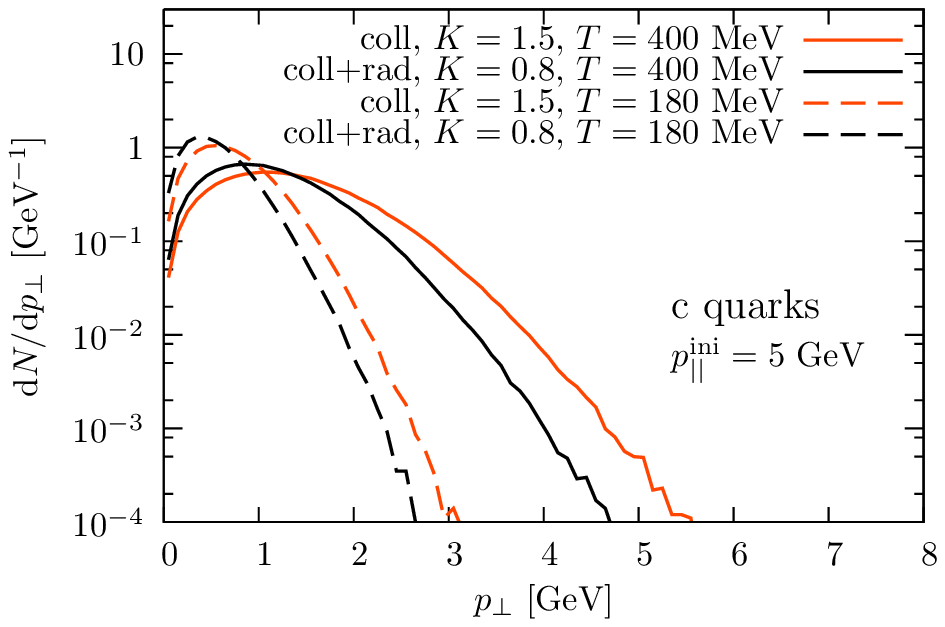}}
 
   \subfigure{\label{fig:evoltimedisperp25}\includegraphics[width=0.48\textwidth]{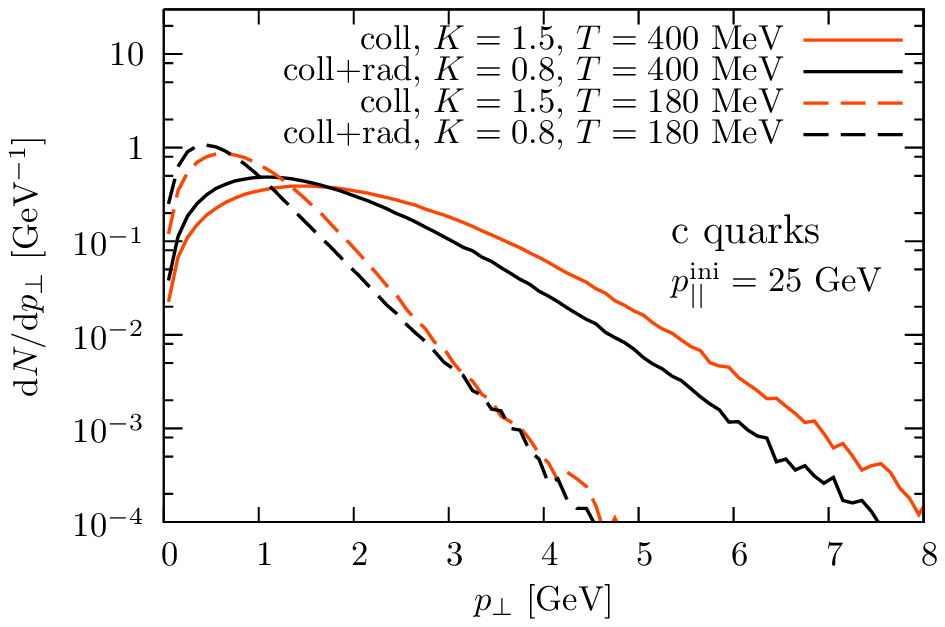}}

 \caption{(Color online) The distribution of $p_\perp$ that a charm quark with $p_{||}^{\rm ini}=5$  \subref{fig:disperp5} and $p_{||}^{\rm ini}=25$  \subref{fig:disperp25} acquires by an explicit propagation through the medium for $\Delta t=1.0$~fm via purely collisional (orange/light) interactions or those including radiative corrections (black/dark) at medium temperature $T=400$~MeV (solid) and $T=180$~MeV (dashed).}
 \label{fig:evoltimedispperp}
 \end{figure}

We now investigate what happens, when the heavy quarks initialized with $\vec{p}^{\,\rm ini}_{||}$ evolve for a small but finite period of time, $\Delta t=1.0$~fm, in a static and infinite medium at temperature $T$. We plot the final $p_\perp$ distribution in figure \ref{fig:evoltimedispperp}.
The distribution is normalized to unity with respect to the integration over $p_\perp$. 
For a decreasing time step and considering only heavy quarks, which acquired a finite $p_\perp>0$, i.e., which underwent a scattering process, one expects that figure \ref{fig:evoltimedispperp} is just the distribution in figure \ref{fig:timedispperp} multiplied by $\Delta t$. For the given time step $\Delta t$ here, this is not observed due to multiple interactions.
One sees that after the propagation of the heavy quarks over this finite period of time the effect of the higher scattering rate in the purely collisional case is even more pronounced than in figure \ref{fig:timedispperp}. Including the time evolution in the medium, we find that the heavy (anti)quark acquires on average a larger $p_\perp$ via a purely collisional interaction than if one includes radiative corrections as well and adapts the $K$-factor accordingly. Of course, the increase of the average $p_\perp$ with larger $p_{||}^{\rm ini}$ and temperature still holds. This indicates that a full treatment of the evolution is necessary to relate to final observables. 

\begin{figure}[tb]
   \subfigure{\label{fig:timeevol5}\includegraphics[width=0.48\textwidth]{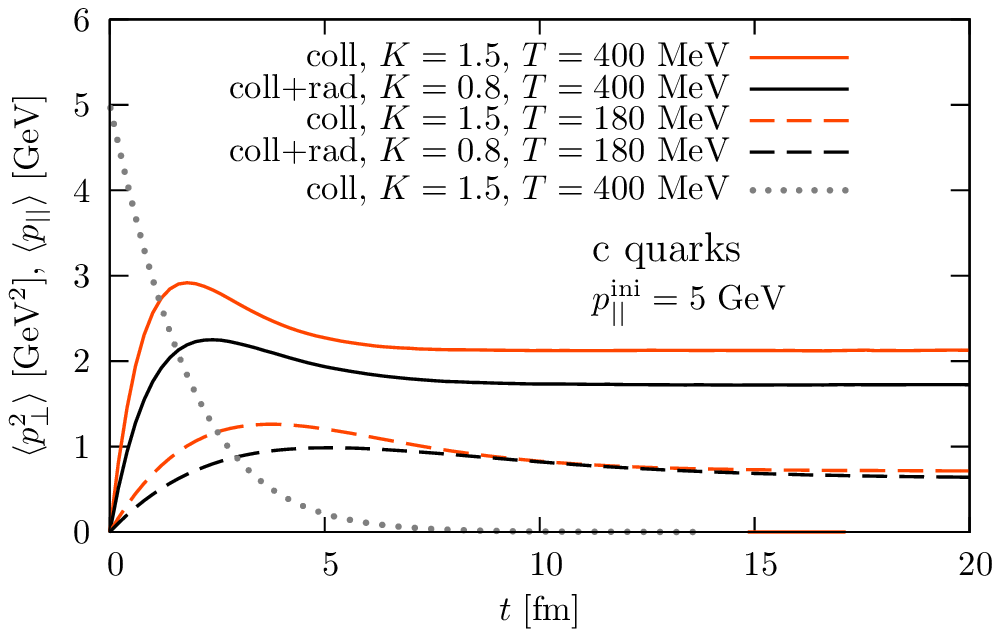}}
 
   \subfigure{\label{fig:timeevol25}\includegraphics[width=0.48\textwidth]{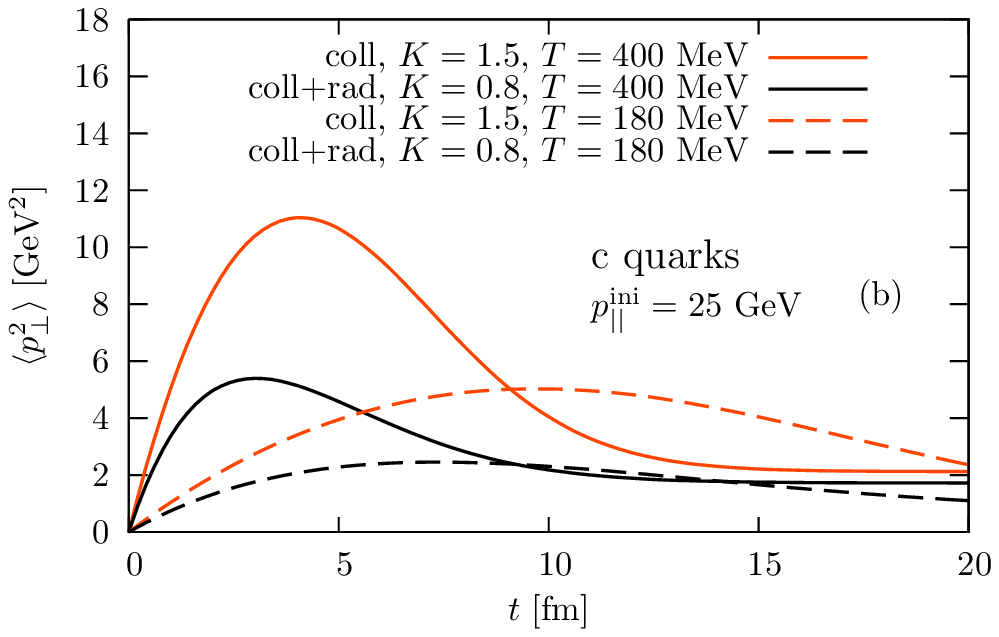}}

 \caption{(Color online) Time evolution of the average $p_\perp^2$ of a charm quark with $p_{||}^{\rm ini}=5$  \subref{fig:timeevol5} and $p_{||}^{\rm ini}=25$  \subref{fig:timeevol25} interacting with the QGP purely collisionally (orange/light) or including radiative corrections (black/dark) at medium temperature $T=400$~MeV (solid) and $T=180$~MeV (dashed). In \subref{fig:timeevol5} we additionally show the time evolution of the average $p_{||}$ for one scenario.}
 \label{fig:timeevol}
 \end{figure}

We can show the same feature by looking at the time evolution of the average $p_\perp^2$ of heavy quarks in a static and infinite medium. This is important because once the heavy quark has undergone a couple of collisions and thus acquired some finite $p_\perp^2$ and lost some of its $p_{||}$ it is already deflected from its initial direction  $\vec{p}^{\,\rm ini}$. At some point the subsequent collisions thus cease to increase the final $p_\perp^2$ with respect to the initial direction. Indeed we observe in figure \ref{fig:timeevol} that after an increase of  $p_\perp^2$ in the beginning of the evolution it decreases again. This is also comprehensible in view of the thermalization occurring on a longer time scale,  when on average $p_{||}=0$.
This is illustrated by the time evolution of the average $p_{||}$ for one interaction mechanism as an example in figure \ref{fig:timeevol5}. The increase of $\langle p_\perp^2\rangle$ stops when $\langle p_{||}\rangle$ has lost already a substantial part ($\sim50$~\%) of its initial value.

We have to note that in principle both types of interaction mechanisms should reach the limit of thermal equilibrium, which is obviously not the case in figure \ref{fig:timeevol}. The reason for this is that for the radiative case the backward mechanism of $3\to2$ processes is not implemented and thus detailed balance is not fulfilled. We think, however, that this does not have significant consequences in the dynamical evolution, which is studied in the following. During the fast expansion the medium cools and dilutes such that the backward reaction is expected to be less effective.

From the study in this section we expect that the clear differences observed in the average $p_\perp^2$ will be relevant for the following investigation of the azimuthal correlations of $Q\bar{Q}$ pairs in heavy-ion collisions.

\section{Azimuthal correlations}\label{sec:correlations}
In this section we investigate the azimuthal correlations of heavy-quark pairs $Q\bar{Q}$ for Pb+Pb collisions at $\sqrt{s}=2.76$~TeV. 

\subsection{Back-to-back initialization}\label{sec:correlations1}

\begin{table*}{
\subtable[\enspace charm]{
\centering
\begin{tabular}{|c|c|c|c|}
  \hline
   & $0-20$~\% & $20-40$~\%  & $40-60$~\%\\
  \hline
  $p_T\in[1-4]$~GeV &  & $0.87$ & $0.735$ \\
  \hline
  $p_T\in[4-10]$~GeV & $0.215$ & $0.14$ &  $0.078$\\  
  \hline
  $p_T\in[10-20]$~GeV & $0.035$ & $0.028$ &  $0.018$\\
  \hline
\end{tabular}
\label{tab:variances1}
}\hfill
\subtable[\enspace bottom]{
\centering
\begin{tabular}{|c|c|c|c|}
  \hline
   & $0-20$~\% & $20-40$~\%  & $40-60$~\%\\
  \hline
  $p_T\in[1-4]$~GeV & $1.22$ & $0.96$ & $0.65$ \\
  \hline
  $p_T\in[4-10]$~GeV & $0.3$ & $0.18$ & $0.095$ \\  
  \hline
  $p_T\in[10-20]$~GeV & $0.038$ & $0.029$ &  $0.019$\\
  \hline
\end{tabular}
\label{tab:variances2}}

\subtable[\enspace charm]{
\centering
\begin{tabular}{|c|c|c|c|}
  \hline
   & $0-20$~\% & $20-40$~\%  & $40-60$~\%\\
  \hline
  $p_T\in[1-4]$~GeV & $1.18$ & $0.84$ & $0.54$ \\
  \hline
  $p_T\in[4-10]$~GeV & $0.13$ & $0.09$ & $0.053$ \\  
  \hline
  $p_T\in[10-20]$~GeV & $0.023$ & $0.019$ &  $0.012$\\
  \hline
\end{tabular}
\label{tab:variances3}}
\hfill
\subtable[\enspace bottom]{
\centering
\begin{tabular}{|c|c|c|c|}
  \hline
   & $0-20$~\% & $20-40$~\%  & $40-60$~\%\\
  \hline
  $p_T\in[1-4]$~GeV & $0.84$ & $0.65$ & $0.42$ \\
  \hline
  $p_T\in[4-10]$~GeV & $0.14$ & $0.093$ & $0.055$ \\  
  \hline
  $p_T\in[10-20]$~GeV & $0.026$ & $0.019$ &  $0.013$\\
  \hline
\end{tabular}
\label{tab:variances4}}
\caption{The variances of the correlation distribution for charm quarks \subref{tab:variances1} and \subref{tab:variances3}, bottom quarks \subref{tab:variances2} and  \subref{tab:variances4} for purely collisional interaction scenarios in \subref{tab:variances1} and \subref{tab:variances2}, and collisional plus radiative corrections in \subref{tab:variances3} and \subref{tab:variances4}.}
\label{tab:variances}
}
\end{table*}

\begin{figure*}[tb]
   \subfigure{\label{fig:cccentral}\includegraphics[width=0.48\textwidth]{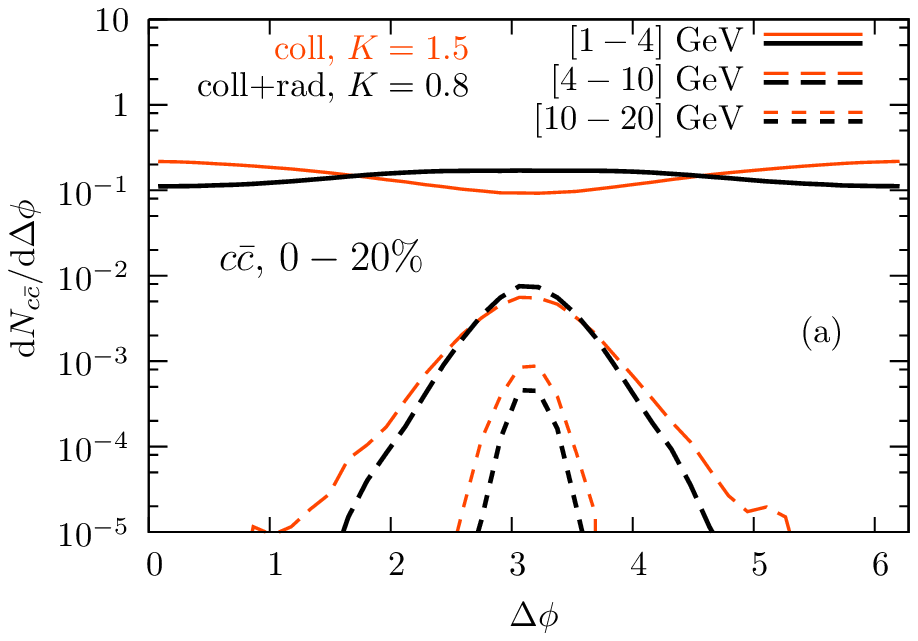}}\hfill
   \subfigure{\label{fig:bbcentral}\includegraphics[width=0.48\textwidth]{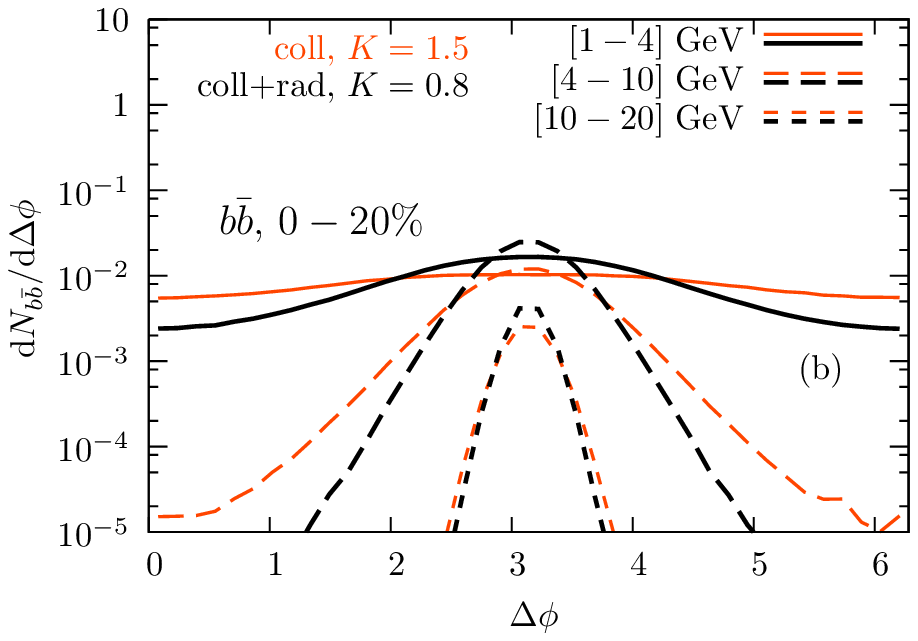}}

   \subfigure{\label{fig:cc2040}\includegraphics[width=0.48\textwidth]{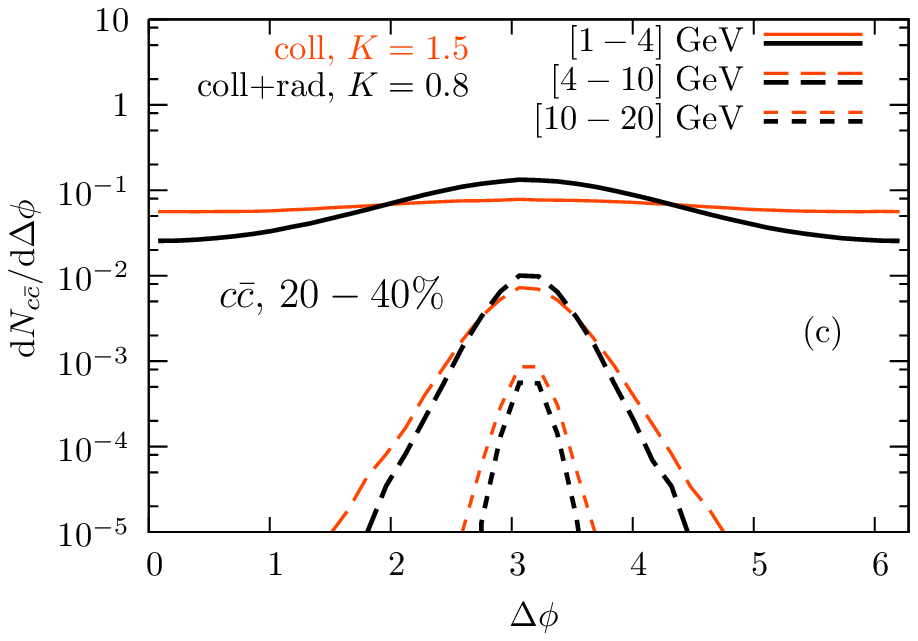}}\hfill
    \subfigure{\label{fig:bb2040}\includegraphics[width=0.48\textwidth]{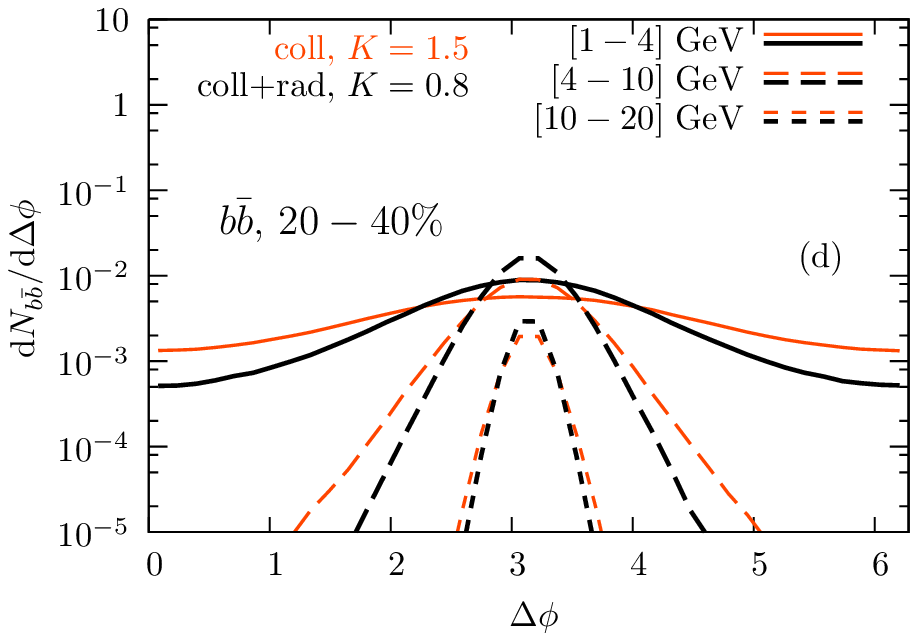}}
 
    \subfigure{\label{fig:cc4060}\includegraphics[width=0.48\textwidth]{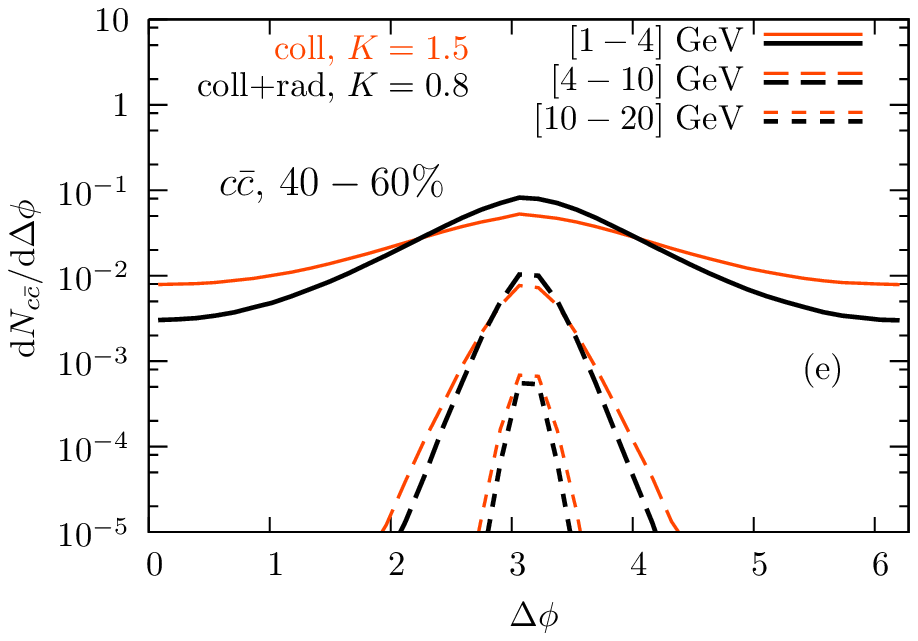}}\hfill
    \subfigure{\label{fig:bbc4060}\includegraphics[width=0.48\textwidth]{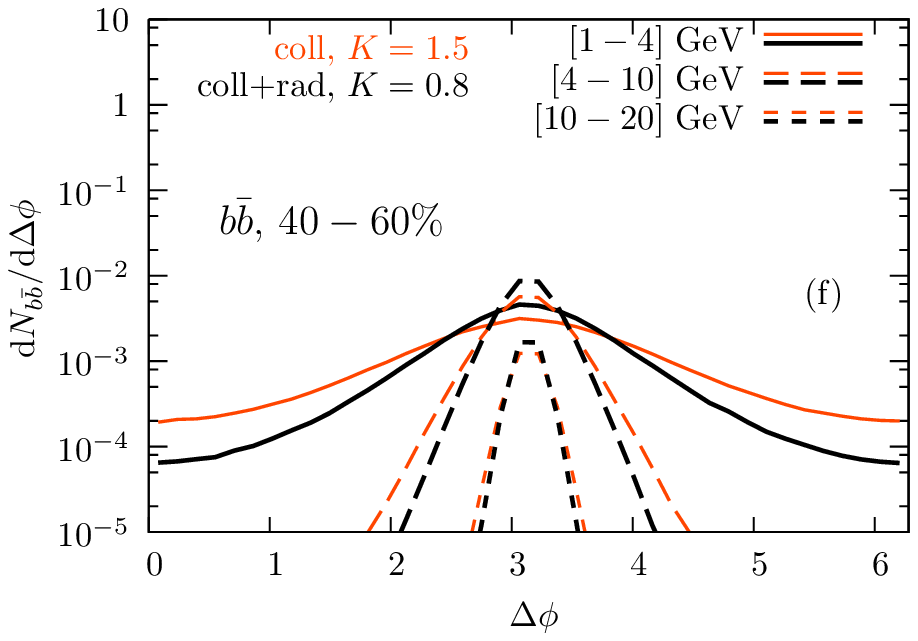}}
 \caption{(Color online) Azimuthal correlations of initially correlated $Q\bar{Q}$ pairs at the transition temperature. In the left column the azimuthal distributions of $c\bar{c}$ pairs are shown, in the right column those of $b\bar{b}$ pairs at midrapidity. The centralities are $0-20$~\% (upper row), $20-40$~\% (middle row) and $40-60$~\% (lower row). In each plot we compare the purely collisional (orange/light) to the collisional plus radiative (black/dark) interaction mechanism for different classes of final $p_T$. See text for more details.}
 \label{fig:azicors}
 \end{figure*}

After initialization of the $Q\bar{Q}$ pairs according to the $p_T$ distribution from FONLL \cite{FONLL1, FONLL2, FONLL3} and the LO production processes, i.e., an azimuthally back-to-back initialization of the $Q\bar{Q}$ pairs with $\vec{p}_{T,\bar{Q}}=-\vec{p}_{T,Q}$, the heavy (anti)quarks are propagated through the QGP by means of the coupled MC@sHQ+EPOS approach, which was described in section \ref{sec:model}. Here, we track the evolution of the heavy (anti)quark until it leaves the QGP. At this transition point we extract the difference of the azimuthal angles, $\Delta\phi$, of those $Q\bar{Q}$ pairs which were initially produced together. The distributions of $\Delta\phi$ are shown in figure \ref{fig:azicors} for $c\bar{c}$ pairs in the left column and for $b\bar{b}$ pairs in the right column. These pairs are taken into account if both the quark and the antiquark are finally at a rapidity $|y_{Q}|<1$ and $|y_{\bar{Q}}|<1$. The results for the $0-20$~\% most central collisions are plotted in the upper row, while in the middle row we see results for $20-40$~\% centrality and in the lowest row for peripheral collisions ($40-60$~\% most central). In each individual plot we show the distribution of azimuthal correlations for three different classes of $p_T$. The lowest $p_T$ class collects all $Q\bar{Q}$ pairs, where both the quark and the antiquark have a final $p_T$ between $1$ and $4$~GeV. In the intermediate-$p_T$ class quark and antiquark have a final $p_T$ between $4$ and $10$~GeV and in the higher $p_T$-class the final $p_T$ of the quark and antiquark is between $10$ and $20$~GeV.

Before we enter into a detailed discussion of the individual plots and the different effects which become apparent in different systems, let us generally comment on one important feature: In all systems and kinematic classes the initial correlations are broadened and they are broadened more strongly for the purely collisional interaction mechanism than for the mechanism including radiative corrections. This can be seen as a direct consequence of the larger average $p_\perp^2$ per unit time for the purely collisional interaction mechanism, as has been shown in the previous section.

\begin{figure}[tb]
 \centering
\includegraphics{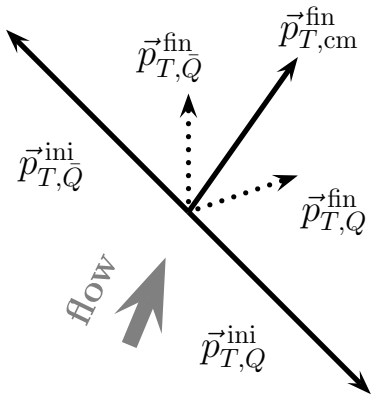}
\caption{Sketch for the definition of quantities used to describe the ``partonic wind'' effect on low-$p_T$ heavy (anti)quarks.  Here, $\vec{p}_{T,{\rm cm}}^{\rm \; fin} =\vec{p}_{T,Q}^{\rm \; fin}+\vec{p}_{T,\bar{Q}}^{\rm \; fin}$.}
\label{fig:sketch}
\end{figure}

The systems that are created in the most central collisions are the largest and reach the highest temperatures and densities. Here, we expect therefore the most efficient broadening of the initial delta-function-like correlations. Indeed, we find a substantial broadening of these correlations for all $p_T$ classes and both interaction mechanisms for $c\bar{c}$ pairs in figure \ref{fig:cccentral} and for $b\bar{b}$ pairs in figure \ref{fig:bbcentral}. Let us first look at the pairs with lowest $p_T$, for which the initial correlations are almost completely washed out. This is a sign of thermalization of the heavy quarks within the QGP. A flat ${\rm d} N_{Q\bar{Q}}/{\rm d}\Delta\phi$ distribution does, however, not imply that the system is necessarily equilibrated, because we do not learn anything about the momentum distribution. Here, we would like to comment on the ``partonic wind'' effect, which was introduced and advocated in \cite{Zhu:2007ne}. It says that the initial back-to-back correlations of $c\bar{c}$ pairs are not only completely washed out but due to the radial flow of the matter the $c\bar{c}$ pairs are pushed into the same direction toward smaller opening angles. Thus, a final enhancement of the azimuthal correlations in the region of $\Delta\phi\simeq0$ is expected. We observe this effect in the lowest $p_T$-class, but only for the purely collisional interaction mechanism. For the mechanism including radiative corrections the broadening of the correlations is not affected by the radial flow in the same manner, and we do not observe a final correlation around $\Delta\phi\simeq0$. 
To quantify this effect we look at the average final center-of-mass transverse momentum of the $c\bar{c}$-pair divided by the average of the sum of the initial transverse momenta $F=\langle |p_{T,{\rm cm}}^{\rm fin}|\rangle/\langle |p_{T,Q}^{\rm ini}|+|p_{T,\bar{Q}}^{\rm ini}|\rangle$, see the sketch in figure \ref{fig:sketch}. Initially, $F$ vanishes. It also vanishes if there is no preferred local direction in the medium. A finite value, on the contrary, indicates the existence of this preferred local direction, here given by the collective flow of the medium. For the lowest $p_T$ class, $F\simeq 0.7$ for the purely collisional interaction mechanism and $F\simeq 0.53$ for the collisional mechanism including radiative corrections.
Due to the larger mass the $b\bar{b}$ pairs are obviously less affected by the ``partonic wind'' effect in accordance with figure \ref{fig:drag}. It also disappears for $c\bar{c}$ pairs in higher $p_T$ classes, where $F\simeq0.1$ for both interaction mechanisms, because the (anti)quarks are too energetic to be substantially affected by the radial flow, and in more peripheral collisions. 

With increasing $p_T$ we see that the peak around $\Delta\phi=\pi$ is less and less broadened. The reason for this is
twofold: a) Particles with larger $p_T$ leave the system more rapidly, so their initial correlation is therefore less affected by the medium, and
b) for asymptotically large initial momenta the time the heavy quarks spend in the medium  is of the order of the radius $R$. The angular opening is thus of the order $\sqrt{\Delta\langle p_\perp^2\rangle}/p_T\sim\sqrt{({\rm d}\langle p_\perp^2\rangle/{\rm d }t)\, R}/p_T$, which is a decreasing function of $p_T$ due to the moderate 
increase of the average $p_\perp^2$ per unit time as a function of $p_{||}^{\rm ini}$ with $p_T$, see figure \ref{fig:pperp2}. By comparing the correlations for $c\bar{c}$ and $b\bar{b}$ pairs, it seems that the 
heavier quarks suffer from larger broadening in the largest $p_T$ class, especially for the most central collisions. 
To understand this fact, one should note that there are two different contributions to each $p_T$ class. First, there are those pairs which were already created in this $p_T$ class 
and do not lose enough energy to end up in a lower $p_T$ class. Second, there are those pairs, which were created at larger $p_T$, but lost approximately the same amount of energy to fall into the respective $p_T$ class. The same considerations as above apply: first, although the scattering rates and the average $p_\perp^2$ per unit time are approximately similar for charm and bottom quarks, see figures 
\ref{fig:Nscat} and \ref{fig:pperp2}, high-$p_T$ bottom quarks stay longer in the QGP than do the charm quarks. The $b\bar{b}$ pairs have thus 
more time to develop a broader correlation peak. Second, bottom quarks lose on average significantly less energy within the medium than do charm quarks, see figure \ref{fig:drag}. Thus, the 
$b\bar{b}$ pairs which are found in a certain $p_T$-trigger class typically have a smaller initial $p_T$ than the equivalent $c\bar{c}$ pairs
and then suffer from larger angular deflections $\propto \sqrt{\Delta\langle p_\perp^2\rangle}/p_T$. This is what we observe in the higher 
$p_T$ classes, where thermalization does not play a role.

The broadening of the correlations can be quantified by looking at the variances of the broadened peak around $\Delta\phi=\pi$. In order to calculate these variances, 
we subtract a background of $Q\bar{Q}$ pairs whose correlations are completely washed out and which we define by the minimum of the angular distribution. After this subtraction we normalize the resulting distribution with respect to the angular integration. The values are given in table \ref{tab:variances} and should be understood with an error of 
$5-10$~\% corresponding to the uncertainties in the subtraction of the background. 
First, we find the confirmation that in the higher $p_T$ classes the peak of the $b\bar{b}$ pairs is broader. Second, the broadening of the correlations is larger for the purely collisional interaction mechanism than for the collisional interaction mechanism including radiative corrections -- for all centralities and $p_T$-trigger classes. The ratio of the variance of the purely collisional over the variance of the collisional plus radiative interaction mechanism is $\simeq 1.5$. This is a direct consequence of the fact that the average $p_\perp^2$ per unit time is larger for the purely collisional interaction mechanism as discussed in section \ref{sec:interactions}.

In an experimental situation it might not always be possible to identify a heavy quark and antiquark as having been initially produced in a pair. 
In particular, there are many $c\bar{c}$ pairs produced in one event. This inability of determining experimentally an initially correlated pair 
would result in an uncorrelated background in addition to the distributions of figure \ref{fig:azicors}. It could possibly be removed by mixed-event techniques.

\subsection{Realistic initial quark -- antiquark distribution}\label{sec:correlations2}

In the previous section the azimuthal correlations of the $Q\bar{Q}$ pairs in heavy-ion collisions at the transition temperature from the QGP to the hadronic phase were investigated under the assumption of an initial back-to-back correlation according to the LO flavor-creation process $q\bar{q} \; (gg)\to Q\bar{Q}$. This is not realistic at high beam energies. Here, NLO processes become important. Especially the gluon splitting process $g\to Q\bar{Q}$ leads to initial correlations in the region of small angular separations. In this part of the work we will use the distributions of $b\bar{b}$ pairs from MC@NLO \cite{Frixione:2003ei,Frixione:2002ik}, which matches NLO QCD matrix elements with a parton shower evolution (HERWIG \cite{Corcella:2000bw,Corcella:2002jc}). This approach is able to reproduce reasonably well the single-inclusive bottom quark distributions and the angular correlations of $b$ jets as measured in $pp$-collisions by the CMS experiment \cite{CMSbbbar}. It had before been tested successful in comparison to $p\bar{p}$ collisions with $\sqrt{s}=2$~TeV at the Tevatron \cite{Frixione:2003ei}. MC@NLO is publically available for bottom quark production but not for charm quark production \cite{mcatnlo}.

\begin{figure}[tb]
  \centering
  \includegraphics[width=0.48\textwidth]{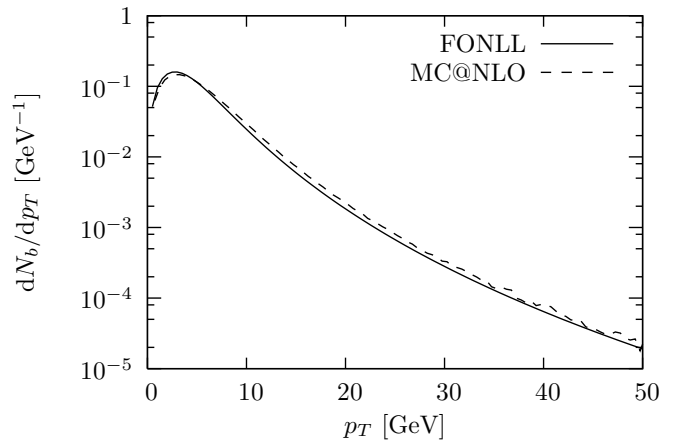}
 \caption{ Comparison of the initial $p_T$ distributions for bottom quarks as obtained from $\sqrt{s}=2.76$~TeV proton-proton collisions in FONLL (solid) and MC@NLO (dashed).}
\label{fig:bbdistr}
\end{figure}

 \begin{figure}[h!]
   \subfigure{\label{fig:bbmcnlo1}\includegraphics[width=0.45\textwidth]{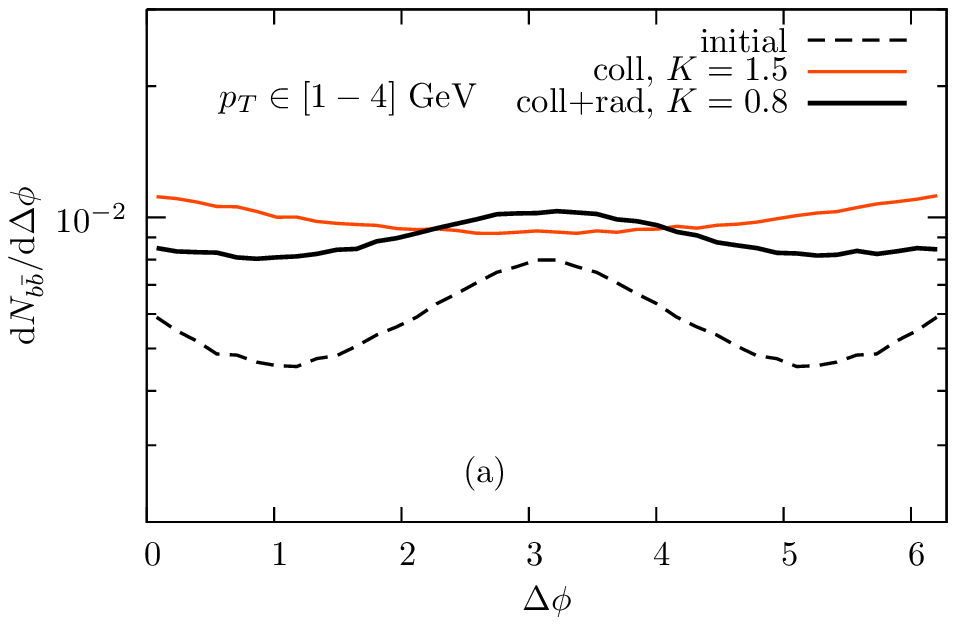}}
   
   \subfigure{\label{fig:bbmcnlo2}\includegraphics[width=0.45\textwidth]{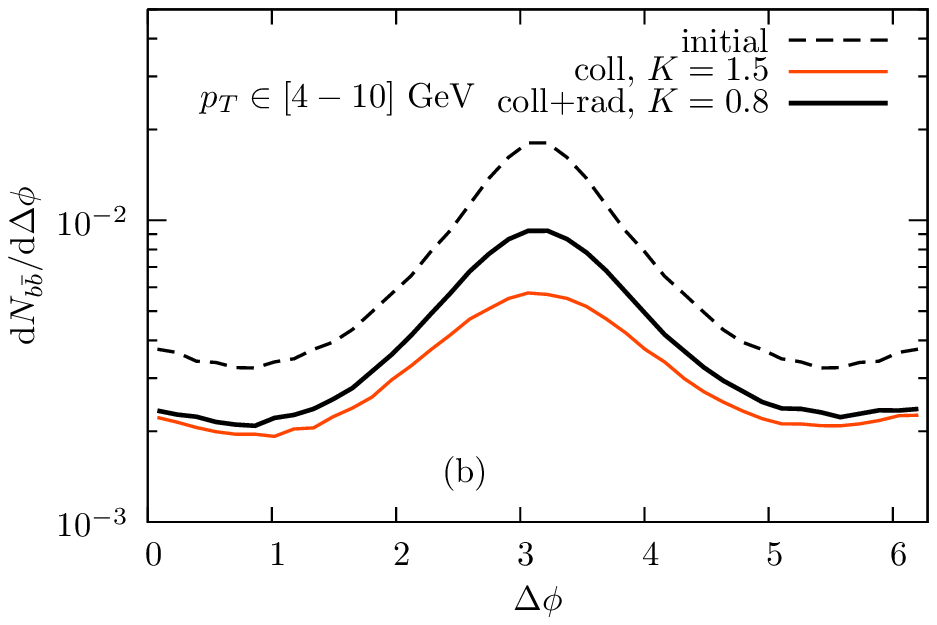}}
 
   \subfigure{\label{fig:bbmcnlo3}\includegraphics[width=0.45\textwidth]{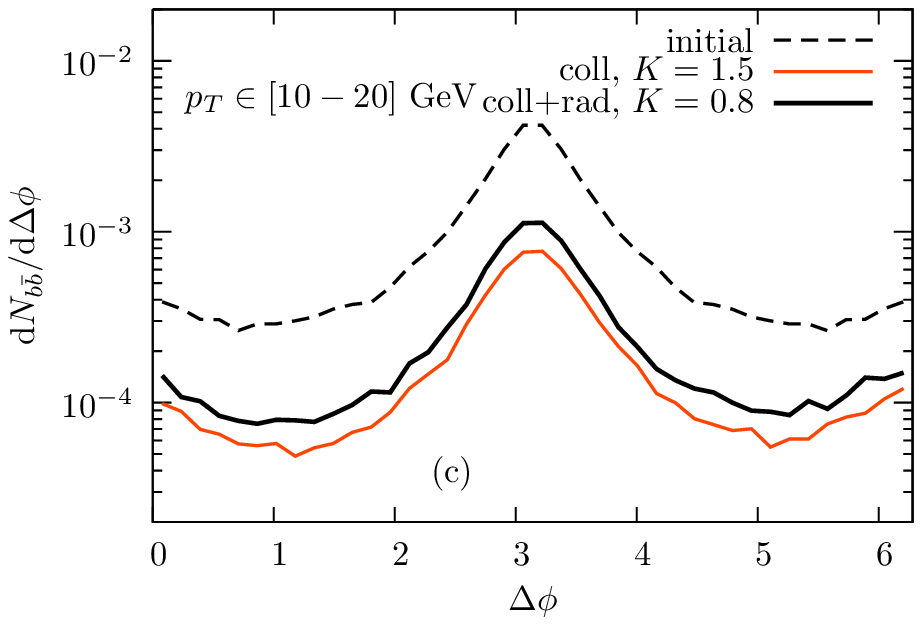}}
 \caption{(Color online) Azimuthal correlations of initially correlated $b\bar{b}$ pairs at the transition temperature from an initialization via MC@NLO. The systems are central Pb+Pb collisions at $\sqrt{s}=2.76$~TeV. The different $p_T$-classes for the final $b$ and $\bar{b}$  are  $[1-4]$~GeV \subref{fig:bbmcnlo1}, $[4-10]$~GeV \subref{fig:bbmcnlo2}, and $[10-20]$~GeV \subref{fig:bbmcnlo3}. In each plot we compare the purely collisional (orange/light) and the collisional+radiative (black/dark) interaction mechanism to the initial distribution (dashed). See text for more details. }
 \label{fig:mcnloazicors}
 \end{figure}

We first compare the initial $p_T$ distributions of bottom quarks at $\sqrt{s}=2.76$~TeV from FONLL, as have been used up to now, to the ones obtained from MC@NLO in figure 
\ref{fig:bbdistr}. We find that the two prescriptions agree well. 
This was also reported in \cite{FONLL3}, where several theoretical models of charm and bottom production at LHC have been compared among each other and versus data. The conclusion reached in that work was that FONLL compares better to data of fully inclusive distributions, which shows the importance of resumming large logarithms of $p_{T,Q}/m_Q$. As mentioned before, within the FONLL more exclusive distributions cannot be calculated however. The difference in the $p_T$ distributions of FONLL from that of MC@NLO is not reflected in the   nuclear modification factor $R_{\rm AA}$ for transverse momenta up to $30$~GeV, which is just as well described starting from either of the distributions.

In the following we use the exclusive $b\bar{b}$ spectra from MC@NLO as an input to the bottom quark propagation in the medium. In figure \ref{fig:mcnloazicors} we show the distributions of the azimuthal correlations of $b\bar{b}$ pairs that are initially correlated for central collisions in the three different 
 $p_T$ classes and for $|y_b|<1$ and $|y_{\bar{b}}|<1$.
Due to the in-medium energy loss at the end of the 
propagation there are fewer $b\bar{b}$ pairs in the higher $p_T$ classes and more in the lower $p_T$ classes than there were initially.  

The initial $\Delta\phi$ distributions of the produced $b\bar{b}$ pairs (dashed curves) have a broadened peak around $\Delta\phi\simeq \pi$ and a second broad and smaller peak at $\Delta\phi\simeq0$. The peak at $\Delta\phi\simeq\pi$ stems from the parton shower evolution of the LO process $q\bar{q}\; (gg)\to b\bar{b}$ and NLO processes $q\bar{q}\; (gg)\to b\bar{b}g$, while the one at $\Delta\phi\simeq0$ is dominated by gluon splitting processes $g\to b\bar{b}g$. We observe that the relative fraction of gluon splitting contributions compared to the back-to-back process is smaller for higher $p_T$ of the bottom (anti)quarks, in accordance with the low-$x$ gluon fragmentation \cite{Catani:1990eg}.

The propagation by either interaction mechanism, purely collisional or collisional and radiative, leads to a decrease of the $\Delta\phi\simeq\pi$ peak and a flattening of the distribution. This is, as in the previous section, stronger for the lower $p_T$ classes and the purely collisional interaction mechanism. In the lowest $p_T$ class in figure \ref{fig:bbmcnlo1} the $b\bar{b}$ pairs have a slight final enhancement of correlations at small angular separation, $\Delta\phi\simeq0$ in the case of the purely collisional interaction. This shape is a remnant of the initial correlations at $\Delta\phi\simeq0$, where the total number is enhanced due to additional pairs whose initial correlations at $\Delta\phi=\pi$ are washed out or whose initial $p_T$ value was larger.
For $p_T$ between $4$ and $10$~GeV in figure \ref{fig:bbmcnlo2} it can clearly be seen that the purely collisional interaction mechanism, like in the previous section, 
leads to a larger broadening of the initial correlation than the collisional and radiative interaction mechanism. 
For larger $p_T$, as in figure \ref{fig:bbmcnlo3}, the difference between the two interaction mechanisms cannot be resolved, along the same lines as it decreased for the evolution with the LO initialization. In addition, the NLO initialization dominates the smearing of the LO peak at $\Delta \phi=\pi$.

\begin{table*}{
\centering
\begin{tabular}{|c|c|c|c|}
  \hline
   & initial (MC@NLO) & coll, $K=1.5$  & coll+rad, $K=0.7$\\
  \hline
  $p_T\in[1-4]$~GeV & $0.53$ &  & $0.61$ \\
  \hline
  $p_T\in[4-10]$~GeV & $0.43$ & $0.48$ &  $0.48$\\  
  \hline
  $p_T\in[10-20]$~GeV & $0.30$ & $0.25$ &  $0.23$\\
  \hline
\end{tabular}}
\caption{The variances of the correlation distributions for bottom quarks corresponding to the curves in figure \ref{fig:mcnloazicors}.}
\label{tab:variancesMCNLO}
\end{table*}

In order to quantify the broadening of the azimuthal correlations we present 
the variances of the peak centered at $\Delta \phi=\pi$ in table \ref{tab:variancesMCNLO}. Here, we again subtract a background by determining the minimum of the distribution on each side of the peak and subtracting the smaller value. The location of these minima also define the central peak. The parts of the distributions beyond the central peak are not included in the calculation of the variances. The thus-obtained distributions around the central peak are normalized with respect to the angular integration.
We see clearly that a naive expectation of simply adding the variance of the initial NLO distribution and the variance of the final distribution from LO initialization does not give the variance of the final distribution from NLO initialization because quenching propagates heavy quarks from a higher $p_T$-trigger class to a lower one.
For the lowest $p_T$ class, the peak at $\Delta \phi=\pi$ disappears completely in a purely collisional interaction scenario, while it is still visible for the interaction mechanism including radiative corrections. Here, the variance is increased from $0.53$ (initial NLO) to $0.61$, corresponding to a $\sim 15\%$ increase. For intermediate $p_T$ one finds a slight increase of the variance from $0.43$ (initial NLO) to $0.48$ in both interaction mechanisms, while in the largest $p_T$-trigger class the variances seem to decrease from the initial distribution. Here, however, the variance is not a reliable criterion anymore since the initial and the final distributions deviate from Gaussian shape. If one looks, instead, at the half width at half maximum (HWHM) after background subtraction the same trend as observed for the back-to-back initialization can be recovered. For intermediate $p_T$ the HWHM increases from $0.59$ initially to $0.71$ for the final distribution in the collisional and radiative scenario ($\sim 20\%$  increase) and to $0.79$ in the purely collisional scenario ($\sim 35\%$  increase). For the largest $p_T$ one can find a small residual broadening for the final distributions with respect to the initial one by looking at HWHM of the isolated peak around $\Delta \phi=\pi$.

\section{Conclusions}
\label{sec:conclusions}
We studied the azimuthal correlations of heavy-flavor quark-antiquark pairs in heavy-ion collisions within a Monte Carlo propagation of heavy quarks, MC@sHQ, coupled 
to a fluid dynamical evolution of the strongly interacting medium coming from EPOS initial conditions. We considered two different interaction mechanisms: a purely collisional one and a 
collisional one including radiative corrections, which are rescaled in order to reproduce the $R_{AA}$ value of D mesons in central Pb-Pb collisions at
LHC. This is a large improvement compared to previous works \cite{Zhu:2006er,Zhu:2007ne,Younus:2013be}, where the diffusion coefficient was considered as a free parameter.

These two interaction mechanisms showed clear differences in a thermal, static medium. We found that the average deflection perpendicular to the initial direction of the heavy 
(anti)quark is significantly larger for the (rescaled) purely collisional interaction. In heavy-ion collisions we were able to show that this translates into a more effective 
broadening of the initial correlations. 

In order to come to this conclusion, we first investigated the broadening of an initial back-to-back correlation according to 
the LO flavor-creation process 
$q\bar{q}\; (gg) \to Q\bar{Q}$ for different centralities and $p_T$ classes. For low-$p_T$ heavy (anti)quarks, in particular charm (anti)quarks are supposed to partially thermalize 
inside the medium. Here, we saw 
that the initial correlations are almost completely washed out. Moreover, for a purely collisional interaction mechanism the low-$p_T$ $c\bar{c}$ pairs 
even show a residual correlation in the region of small angular separation. This so-called ``partonic wind'' effect is, however, absent for an interaction mechanism which includes 
radiative corrections.

In the intermediate-$p_T$ region the broadening of the initial correlations was found to be most visible as the heavy (anti)quarks do not thermalize and spend enough time in the medium 
to be significantly affected. Here, we clearly observed that the azimuthal correlations of $Q\bar{Q}$ pairs are broadened more effectively by purely collisional interactions. 
Including radiative corrections we found that the initial azimuthal correlations survive the propagation of the $Q\bar{Q}$ pairs to a larger degree.

Beyond $p_T\sim 10~{\rm GeV/c}$, we found that the broadening due to the interactions in the medium is weaker and likely to be hidden by the NLO corrections
affecting the initialization of the $b\bar{b}$-pairs.

Observables of azimuthal correlations of $Q\bar{Q}$ pairs are thus sensitive to the properties of the interaction of the heavy quarks with the medium and thus of the energy loss 
mechanism. In particular the potential to discriminate between purely collisional interactions and those including radiative corrections makes heavy quark correlations a promising 
supplement to traditional heavy-quark observables such as the nuclear modification factor $R_{\rm AA}$ and the elliptic flow $v_2$. 

As the LHC data allows experimentalists to thoroughly analyze the azimuthal correlations of heavy quarks future work including hadronization and decay channels will provide realistic predictions.

\section*{Acknowledgements}
The authors thank Sarah Lapointe, Andre Mischke, Steffen Bass and Shanshan Cao for fruitful discussions and Stefano Frixione for explanations on MC@NLO. This work was partially supported by the Hessian LOEWE initiative Helmholtz International 
Center for FAIR, the ANR research program ``hadrons @ LHC''  (grant ANR-08-BLAN-0093-02), I3-Hadronphysics, Project TOGETHER (Pays de la Loire) and I3-Hadronphysics II.

\end{document}